\documentclass[a4paper,11pt]{article}
\pdfoutput=1
\usepackage{jheppub}
\usepackage{amsmath,amssymb,bm,graphicx,bbold,epsf,colordvi}
\usepackage{lipsum}
\usepackage{soul}
\usepackage{braket}
\usepackage{bm}
\allowdisplaybreaks 
\usepackage{color}
\usepackage[dvipsnames]{xcolor}

\usepackage{mathrsfs}
\allowdisplaybreaks

\newcommand{\bef}{\begin{figure}[hbt]\centering}
\newcommand{\eef}{\end{figure}}
\usepackage{mathtools}
\usepackage{subfigure}
\usepackage{booktabs}
\usepackage{graphicx}
\graphicspath{ {./figure/} }
\usepackage{caption}
\usepackage{lipsum}

\newcommand{\nnu}{\nonumber\\}

\newcommand{\beq}{\begin{equation}}
\newcommand{\eeq}{\end{equation}}
\def\bea#1\eea{\begin{align}#1\end{align}}

\def \be  {\begin{equation}}
\def \ee  {\end{equation}}
\def \ba  {\begin{eqnarray}}
\def \ea  {\end{eqnarray}}

\allowdisplaybreaks
\newcommand{\cSoft}{{\mathscr S}}

\title{QCD resummation on single hadron transverse momentum distribution with the thrust axis}

\author[a,b,c]{Zhong-Bo Kang,}
\author[a,b,c]{Ding Yu Shao}
\author[a,b]{and Fanyi Zhao}

\affiliation[a]{Department of Physics and Astronomy, University of California, Los Angeles, CA 90095, USA}
\affiliation[b]{Mani L. Bhaumik Institute for Theoretical Physics, University of California, Los Angeles, CA 90095, USA}
\affiliation[c]{Center for Frontiers in Nuclear Science, Stony Brook University, Stony Brook, NY 11794, USA}
                                   
\emailAdd{zkang@physics.ucla.edu}
\emailAdd{dingyu.shao@physics.ucla.edu}
\emailAdd{fanyizhao@physics.ucla.edu}

\abstract{We derive the transverse momentum dependent (TMD) factorization and resummation formula of the unpolarized transverse momentum distribution ($j_T$) for the single hadron production with the thrust axis in  electron-positron collision. Two different kinematic regions are considered, including small transverse momentum limit $j_T \ll Q$, and joint transverse momentum and threshold limit $j_T \ll Q(1-z_h) \ll Q$, where $Q$ and $z_h$ are the hard scattering energy and the observed hadron momentum fraction. Using effective theory methods, we resum logarithms $\ln(Q/j_T)$ and $\ln(1-z_h)$ to all orders. In the end we present the differential cross sections and Gaussian widths calculated for the inclusive charged pion production and find that our results are consistent with the measurements reported by the Belle collaboration. }
\begin{document} 
\maketitle

\section{Introduction}

The transverse momentum dependent parton distribution functions (TMD PDFs) and fragmentation functions (TMD FFs) are the fundamental objects to understand the intrinsic hadron structure, in particular, the three-dimensional (3D) imaging of the hadrons in the momentum space~\cite{Accardi:2012qut,Boer:2011fh,Aidala:2020mzt}. A lot of progress has been made in understanding 3D imagining of the nucleon via both unpolarized and polarized TMD PDFs, see some recent work in Refs.~\cite{Liu:2020dct,Cammarota:2020qcw,Bacchetta:2020gko,Bacchetta:2019sam,Scimemi:2019cmh,Liu:2018trl,Chien:2020hzh,Chien:2019gyf,Fleming:2019pzj,Buffing:2018ggv,Echevarria:2014xaa,Boer:2015vso}. On the other hand, in comparison with the fruitful results in TMD PDFs, the research on the TMD FFs of hadrons definitely needs more development. The current main channels to probe TMD FFs are either semi-inclusive processes in deep inelastic scattering (SIDIS) and in $e^+e^-$ collisions (hadron pair production)~\cite{Bacchetta:2006tn,Kang:2015msa,Echevarria:2020qjk,Callos:2020qtu}, or hadron distribution inside jets~\cite{Arratia:2020nxw,Neill:2016vbi,Kang:2017btw,Kang:2017glf,Kang:2019ahe,Kang:2020xyq,Gutierrez-Reyes:2019msa}. For a recent review on fragmentation functions, see Ref.~\cite{Metz:2016swz}. It will be interesting and instructive to find more observables to probe TMD FFs. 

The extraction of TMD PDFs and/or TMD FFs relies on the so-called TMD factorization in QCD. For example, transverse momentum distribution of the Drell-Yan type processes has been developed in the seminal literature by Collins, Soper and Sterman~\cite{Collins:1984kg} for a long time and is usually referred to as CSS formalism. For a modern reformulation of the CSS formalism, see~\cite{Collins:2011zzd}. Similar formulas describe the TMD factorization for the SIDIS process in electron-nucleon collisions, $e^-p\to e^-hX$~\cite{Ji:2004wu,Collins:1992kk,Boer:2003cm,Bacchetta:2006tn}, and for back-to-back hadron pair production in $e^+e^-$ annihilation, $e^+e^-\to h_1 h_2 X$~\cite{Collins:1981uk,Boer:1997mf,Pitonyak:2013dsu}. The universality for the non-perturbative parametrization has been investigated in~\cite{Cammarota:2020qcw,Bacchetta:2020gko,su:2014wpa,Boglione:2016bph,Hautmann:2020cyp,Collins:2004nx,Kang:2015msa,Bacchetta:2017gcc}. Recently, the TMD factorization structure has been re-investigated using Soft-Collinear Effective Theory (SCET)~\cite{Bauer:2000yr,Bauer:2001yt, Bauer:2002nz, Beneke:2002ph} and the renormalization group (RG) techniques~\cite{Becher:2010tm,Chiu:2011qc,GarciaEchevarria:2011rb}.

The electron-positron collider provides a clean environment to study TMD FFs using the inclusive hadron production, since there is no hadronic contamination from the initial states. The standard process to probe TMD FFs in $e^+e^-$ collisions is the aforementioned back-to-back hadron pair production, $e^+e^-\to h_1 h_2 X$, which probes the same TMD FFs as those in SIDIS process, $e^-p\to e^-hX$~\footnote{Note that the modern formulation of so-called properly-defined TMD FFs combine the usual TMD FFs and the soft function for the process. Here we are referring to the properly-defined TMD FFs. For details, see~\cite{Collins:2011zzd}.}. Recently, the single-hadron differential cross section for the process, $e^+e^-\rightarrow hX$, is reported by the Belle collaboration~\cite{Seidl:2019jei}, where the hadron cross section is studied as a function of the event-shape variable called thrust $T$, fractional energy $z_h$, and the transverse momentum $j_T$ with respect to the thrust axis. The $j_T$ distribution shows the usual Gaussian shape and this gives the hope that through such a new measurement, one would gain better understanding of the same TMD FFs. With such an assumption, some phenomenological work has been performed in~\cite{Boglione:2017jlh,Soleymaninia:2019jqo}. 

In this paper we perform a detailed theoretical study for the Belle observable and we develop a TMD factorization formalism for describing such a $j_T$ distribution. The plane perpendicular to the thrust axis splits the full phase space into two hemispheres. One only measures the hadron $j_T$ in one hemisphere, and the other hemisphere is unmeasured. Such type of measurements are termed as non-global observables~\cite{Dasgupta:2001sh}, which are sensitive to radiation in only a part of phase space. The factorization and resummation formula for non-global observables have a very different structure from the standard global observable~\cite{Sterman:2004en}. For example, the leading-order evolution equation for Non-Global Logarithms (NGLs) is named as BMS equation~\cite{Banfi:2002hw}, which is a non-linear evolution equation. The TMD factorization formalism for the non-global hemisphere event shape has been studied by one of the authors in~\cite{Becher:2017nof}, where they find that the rapidity logarithms evolution does not constitute an essential complicated structure, since it is tied with a universal transverse momentum dependent jet function which also appears in the global observables. Besides, after comparing with the data at the LEP, they also find that the leading non-perturbative effects are related to the Collins-Soper kernel\footnote{In~\cite{Becher:2017nof} the Collins-Soper kernel is named as the collinear anomaly function.}.

We mainly consider the kinematic region with $j_T\ll Q$, where a TMD factorization can be developed which resums $\ln(Q/j_T)$. Here $Q$ is the virtuality of the intermediate photon in $e^+e^- \to \gamma^*$. In this region, Belle collaboration finds that the cross sections can be well described by Gaussians in $j_T$, and that the width of the Gaussians shows an initially rising,
then decreasing $z_h$-dependence when $z_h\to 1$. Because of this, we further consider the threshold $\ln(1-z_h)$ resummation in the $z_h\to 1$ limit. We apply SCET to develop a TMD factorization formalism. Using renormalization group evolution techniques, we resum logarithmic terms to next-to-leading logarithmic (NLL) accuracy, including NGLs. The experimental data are shown as comparison and in good agreement with our theoretical predictions.

The remainder of this paper is organized as follows. In Sec.~\ref{sec:global}, we present a factorized framework, which only resums the so-called global logarithms. This section would allow us to develop intuition for our framework and understand connection to the standard TMD FFs. In Sec.~\ref{sec:full}, we present the full factorization formalism, which allows us to resum both global and NGLs. In Sec. \ref{sec:pheno}, numerical results of differential cross sections for pion production in $e^+e^-\rightarrow \pi^{\pm} X$ are presented, as a function of energy fraction $z_h$ and transverse momentum $j_T$. We also present the Gaussian width for the $j_T$ distribution as computed from our theoretical formalism, and compare them with the Belle experimental data. Finally, conclusions are given in Sec.~\ref{sec:summary}. \\

\section{TMD formalism: global structure} 
\label{sec:global}
We consider the process, $e^+ + e^- \to h\,(z_h,\, j_T)+X$, in $e^+e^-$ annihilation. The center-of-mass (CM) energy of the $e^+e^-$ collisions is given by $s = Q^2 = (p_{e^+} + p_{e^-})^2$, and the hadron momentum fraction $z_h = 2p_h\cdot q/Q^2 = 2E_h/Q$ is measured. In addition, the hadron's transverse momentum $j_T$ is measured with respect to the so-called thrust axis $\hat n$, which maximizes the event-shape variable thrust $T$~\cite{Brandt:1964sa}:
\bea
T \equiv {\rm max}_{\hat n} \frac{\sum_i |\vec{p}_i\cdot \hat n|}{\sum_i |\vec{p}_i|}\,,
\eea
with the momenta $\vec{p}_i$ of the particles measured in the $e^+e^-$ CM frame. For convenience, we align the thrust axis to be along $+z$-direction, and define light-like vectors $n^\mu = (1, \hat n) = (1, 0, 0, 1)$ and $\bar n^\mu = (1, -\hat n) = (1, 0, 0, -1)$. We expand any momentum $p^\mu$ in the light cone frame as $p^\mu=(p^+, p^-, p_T)$ with $p^+ = n\cdot p = p_0 - p_z$ and $p^- = \bar n\cdot p = p_0 + p_z$. It is important to emphasize that even though we measure the hadron transverse momentum $j_T$ with respect to the thrust axis, our cross section is not differential in the thrust variable~$T$. In other words, we consider the cross section for the hadron production, which is differential only in $z_h$~and~$j_T$:
\bea
\frac{d\sigma}{dz_h\,d^2\vec{j}_T}\,.
\label{eq:obs}
\eea
That is to say, the  only purpose of the thrust measurement is to provide the thrust axis $\hat n$ and we sum over the entire thrust region $0.5 < T < 1$. For the cross section that is further differential in the thrust variable $T$, see e.g. Refs.~\cite{Jain:2011iu,Lustermans:2016nvk}. We find that such an observable in Eq.~\eqref{eq:obs} has a better connection to the standard TMD FFs. 

The plane perpendicular to the thrust axis divides the full space into two hemispheres: the one on the right (along $+z$-side) is referred to as the right hemisphere, while the one on the left is the left hemisphere. Note that the observed hadrons are always measured in the right hemisphere, while no measurement is performed for the left hemisphere. Because of this, the kinematics in the left hemisphere are unconstrained, our observable in Eq.~\eqref{eq:obs} is a \textit{non-global observable}~\cite{Dasgupta:2002bw}. Such observables will involve non-global structures which can not be captured by the traditional exponential formula~\cite{Sterman:2004en}. In this paper we will apply the jet effective theory~\cite{Becher:2016omr,Becher:2016mmh} to derive the factorization and resummation formula. Since the full factorization structure is quite complicated which we save for the next section, in this section, we will for the moment ignore the NGLs that arise from the non-global structure, and write down a factorized formalism to resum the global logarithm and build our intuition. 

\begin{figure}[t]
\begin{center}
  \includegraphics[width=0.9\textwidth]{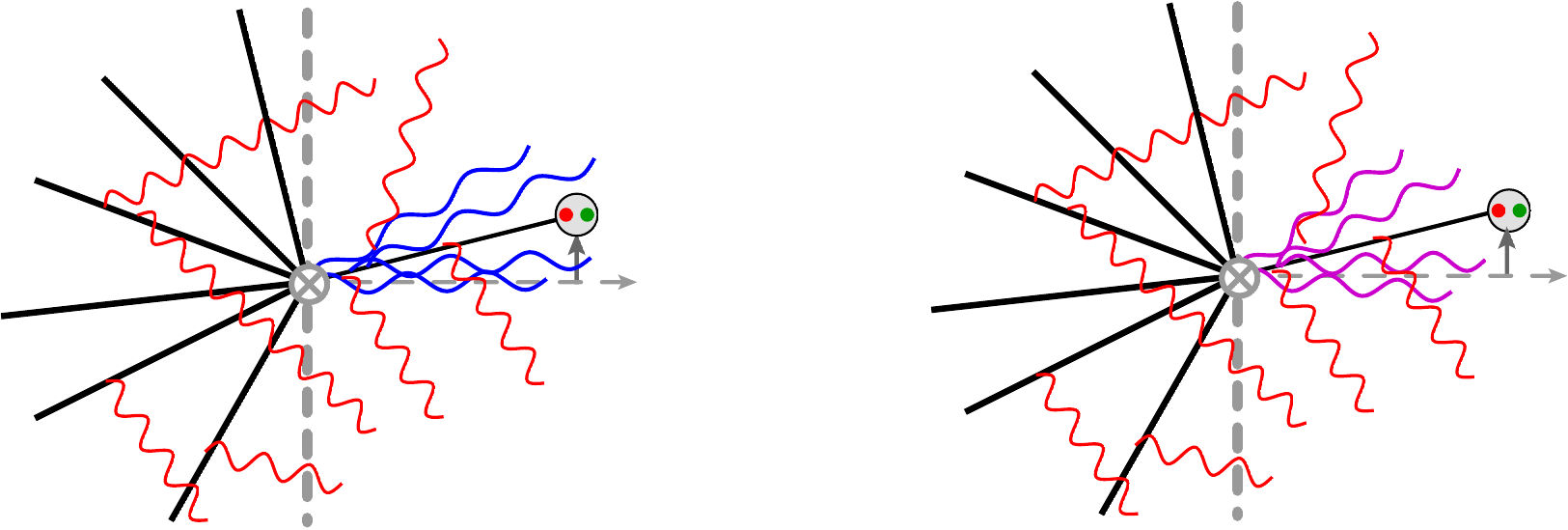}
\end{center}
  \caption{ Hadron transverse momentum $\vec j_T$ with the thrust axis $\hat n$ in two different regions. The black lines represent the energetic partons in the unmeasured {\it left} hemisphere, while the hadron is measured in the {\it right} hemisphere. Vertical dashed line represents a plane that is perpendicular to the thrust axis and that divides the space to left and right hemisphere. The red curves denote soft radiations from the energetic partons with the virtuality of $j_T$. The blue lines in the left panel describe collinear radiations along the thrust axis, while the purple ones in the right panel give collinear-soft (c-soft) radiations . }
\label{fig:eft}
\end{figure} 

\subsection{TMD factorization formalism}
We consider the kinematic region where the transverse momentum is small $j_T \ll Q$, and thus is sensitive to TMD physics. Setting the usual power expansion parameter $\lambda = j_T/Q$, we find that the relevant momentum modes in this region are given by
\begin{itemize}
    \item{\makebox[2cm]{\textbf{hard}:\hfill} $p_h \sim Q(1,1,1)$}
    \item{\makebox[2cm]{{\color{blue}\textbf{collinear}}:\hfill} $ p_{c} \sim Q(\lambda^2, 1, \lambda)$}
    \item{\makebox[2cm]{{\color{red}\textbf{soft}}:\hfill} $ p_s \sim Q(\lambda, \lambda, \lambda)$}
\end{itemize}
The different modes are illustrated in Fig.~\ref{fig:eft} (left). The hard modes encode energetic radiations in the {\it left} hemisphere: since the hadron is observed in the right hemisphere and has $j_T\ll Q$, any energetic radiation in the right hemisphere will lead to a large transverse momentum for the hadron and thus move the hadron out of the kinematic $j_T\ll Q$ region; consequently such radiation is not allowed in the right hemisphere. On the other hand, soft and collinear modes have the same transverse momentum of $j_T$, and thus both contribute to our observable. The difference is that collinear modes encode energetic radiations along the thrust axis, while soft modes describes large angle long wave radiations. Based on the mode analysis, the factorization formalism is given as
\begin{align}
\frac{d\sigma}{dz_hd^2\vec{j}_T} =&\, \sigma_0 \sum_{i = q, \bar q, g} e_q^2 \int d^2{\vec k}_T\, d^2{\vec \lambda}_T\, \delta^{(2)}\left(\vec j_T - \vec k_T - z_h \vec \lambda_T \right) 
\nonumber \\
&\times \mathcal{H}^i(Q, \mu) D_{h/i}(z_h, k_T, \mu, \nu) \mathcal{S}_i(\lambda_T, \mu, \nu)\,,
\label{eq:global}
\end{align}
where $D_{h/i}(z_h, k_T, \mu, \nu)$ is the usual TMD FF with $k_T$ the transverse momentum of the hadron $h$ with respect to the fragmenting parton $i$. On the other hand, $\mathcal{S}_i(\lambda_T, \mu, \nu)$ is the soft function, with $\mu$ and $\nu$ renormalization and rapidity scales, respectively. The leading-order (LO) cross section is given by
\bea
\sigma_0 = \frac{4\pi\alpha_{\rm em}^2}{3Q^2}\,,
\eea
with $\alpha_{\rm em}$ the fine structure constant. Note that the factorization in Eq.~\eqref{eq:global} neglects the power corrections from the ratios $j_T^2/Q^2$, which is small in the kinematic $j_T\ll Q$ region we consider. Nevertheless, in the region of $j_T\sim Q$ one can include such power corrections from the fixed-order calculations~\cite{Moffat:2019pci}. This is usually referred to as the $Y$-term in the CSS formalism~\cite{Collins:1984kg,Collins:2011zzd}. 
 
\begin{figure}[htb]
\begin{center}
  \includegraphics[width=0.9\textwidth]{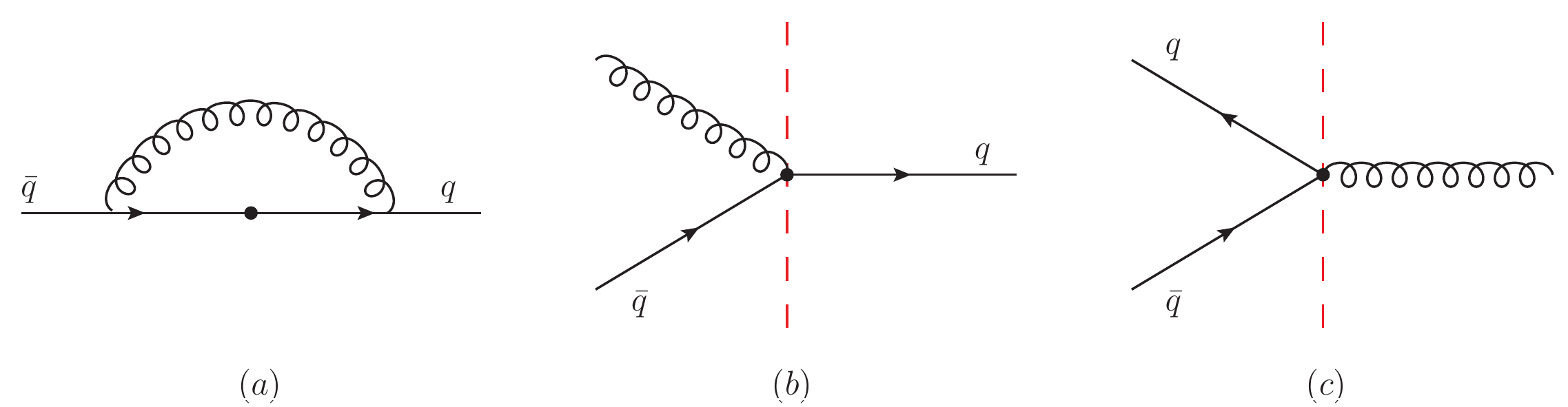}
\end{center}
  \caption{Three configurations that contribute to the hard function: (a) virtual correction; (b) quark $q$ is on the right hemisphere, while both anti-quark $\bar q$ and gluon $g$ are on the left hemisphere; (c) gluon $g$ is on the right hemisphere, while both quark $q$ and anti-quark $\bar q$ are on the left hemisphere. Note that the observed hadron is on the right hemisphere.}
\label{fig:configs}
\end{figure} 

It is important to emphasize that the above TMD formalism is already different from the earlier conjectures used in~\cite{Boglione:2017jlh,Soleymaninia:2019jqo}. In particular, at leading power, our formalism depends on both quark and gluon TMD FFs, while the previous conjecture contains only quark TMD FFs. To convince that this has to be the case, the easiest way is to look at the Feynman diagram configurations that contribute to our observable at the next-to-leading order (NLO), from which we also derive the hard functions $H^i$ with $i=q\,(\bar q),\, g$. At LO, we produce back-to-back quark $q$ and anti-quark $\bar q$, each in their corresponding left or right hemisphere, and our hard function is normalized to be $H = 1$ at this order. At NLO, we receive three contributions as shown in Fig.~\ref{fig:configs}. Here, Fig.~\ref{fig:configs} (a) is the virtual correction to the LO process $e^+e^- \to q\bar q$ and $q$ (or $\bar q$) later on fragments into the observed hadron $h$, and thus this contribution is associated with the quark TMD FFs $D_{h/q}(z_h, k_T, \mu, \nu)$ in Eq.~\eqref{eq:global}. Fig.~\ref{fig:configs} (b) and (c) describe the hard scattering with three partons in the final state, where two hard partons are emitted in the left hemisphere and one parton $i$ in the right hemisphere. Here, for three-particle final states the thrust axis $\hat n$ is determined by the direction of the most energetic parton. For (b), it is $\bar q$ and $g$ on the left hemisphere, while $q$ on the right hemisphere, which fragments into the hadron $h$ and thus we have quark TMD FF $D_{h/q}$. For (c), it is $q$ and $\bar q$ on the left hemisphere, while $g$ on the right hemisphere which fragments into the hadron $h$ and thus we have gluon TMD FF $D_{h/g}$ in Eq.~\eqref{eq:global}. We emphasize again that no hard radiations are allowed in the right hemisphere to maintain $j_T\ll Q$. 

Direct calculations give us the following expressions for the corresponding bare hard functions at NLO, 
\begin{subequations}
\bea
&\mathcal{H}^{q}_2 =\mathcal{H}^{\bar q}_2 \equiv \mathcal{H}_{(a)}^q(Q,\epsilon) =  1 + \frac{\alpha_s}{4\pi} C_F \left( \frac{\mu^2}{Q^2} \right)^{\epsilon} \left[ - \frac{4}{\epsilon^2} - \frac{6}{\epsilon} -16 + \frac{7}{3}\pi^2\right]\,,
\\
&\mathcal{H}_3^{q} = \mathcal{H}_3^{\bar q}\equiv \mathcal{H}_{(b)}^q(Q,\epsilon) =   \frac{\alpha_s}{4\pi} C_F \left( \frac{\mu^2}{Q^2} \right)^{\epsilon} \left[ \frac{2}{\epsilon^{2}}+\frac{3}{\epsilon}+\frac{29}{3}-\frac{3 \pi^{2}}{2}-2 \ln^{2} \left(2\right)\right.
\nonumber \\
& \hspace{4cm}\left.+\frac{5 \ln \left(3\right)}{4}-4 \operatorname{Li}_{2}\left(-\frac{1}{2}\right)\right]\,,
\\
&\mathcal{H}_3^{g} \equiv \mathcal{H}_{(c)}^g(Q,\epsilon) =  \frac{\alpha_s}{4\pi} C_F \left[ -\frac{1}{6}+\frac{\pi^{2}}{3}+2 \ln^{2}\left(2\right)-\frac{5 \ln \left(3\right)}{4}+4 \operatorname{Li}_{2}\left(-\frac{1}{2}\right) \right]\,,
\eea
\end{subequations}
where we use the notation $\mathcal{H}^{i}_{m}$ with index $m=2,3$ at NLO representing the number of final-state partons, while the subscripts $(a)$, $(b)$, $(c)$ correspond to the configurations in Fig.~\ref{fig:configs}. We include the LO result into $\mathcal{H}^{i}_2$, and we note that $\mathcal{H}_3^{g}$ starts at $\mathcal{O}(\alpha_s)$ order, which is free of any divergence. Note that the function $\mathcal{H}^{q}_{2}$ is the standard dijet hard function, that arise in e.g. back-to-back hadron pair production~\cite{Kang:2015msa}. If one ignores the non-global structure, the renormalization group (RG) equation for the hard function can be easily obtained from the above expressions. However, the structure for the full RG equations is much more complicated and will be shown in the next section. 

\subsection{TMD formalism in coordinate space}
TMD formalism in Eq.~\eqref{eq:global} involves convolution over the transverse momentum $\vec{k}_T$ and $\vec{\lambda}_T$. We apply the Fourier transform to go into the coordinate $b$-space and thus the convolution becomes a simple product. To get started, realizing
\bea
\delta^{(2)}\left(\vec j_T - \vec k_T - z_h \vec \lambda_T \right) = \frac{1}{z_h^2}\int \frac{d^2\vec{b}}{(2\pi)^2} e^{i\vec{b}\cdot\left(\vec{j}_T/z_h - \vec{k}_T/z_h - \vec{\lambda}_T\right)}\,,
\eea
and we thus can write Eq.~\eqref{eq:global} in the following form
\begin{align}
\frac{d\sigma}{dz_hd^2\vec{j}_T} =\, \sigma_0 \sum_{i = q, \bar q, g} e_q^2 \int \frac{d^2\vec{b}}{(2\pi)^2} 
e^{i\vec{b}\cdot \vec{j}_T/z_h} \mathcal{H}^i(Q, \mu)D_{h/i}(z_h, b, \mu, \nu) \mathcal{S}_i(b, \mu, \nu)\,,
\label{eq:global-b}
\end{align}
where the $b$-space TMD FF and soft function are defined as 
\bea
D_{h/i}(z_h, b, \mu, \nu)  =&\, \frac{1}{z_h^2}\int d^2\vec{k}_T e^{-i\vec b\cdot \vec k_T/z_h} D_{h/i}(z_h, k_T, \mu, \nu)\,,
\\
\mathcal{S}_i(b, \mu, \nu) = & \int d^2\vec{\lambda}_T e^{-i\vec b\cdot \vec \lambda_T} \mathcal{S}_i(\lambda_T, \mu, \nu)\,.
\eea
Both TMD FFs and soft function suffer from rapidity divergence, which was regularized via the rapidity regulator in~\cite{Chiu:2011qc,Chiu:2012ir}. As a consequence we have rapidity poles in $1/\eta$ and the associated rapidity scale $\nu$, besides the usual poles in $1/\epsilon$ in dimensional regularization and the associated renormalization scale $\mu$. In order to resum relevant logarithms, one can use transitional CSS formalism~\cite{Collins:1984kg}, or effective theory approaches \cite{Becher:2010tm,Chiu:2011qc,Chiu:2012ir}~\footnote{We recommend \cite{Ebert:2019okf} for the comparison for different TMD factorization frameworks.}. The NLO perturbative expressions for TMD FFs are well-known~\cite{Kang:2017glf}, and we list here for completeness:
\begin{subequations}
\label{eq:TMDb}
\bea
\label{eq:qq}
D_{q/q}(z_h, b, \mu, \nu) =& \frac{1}{z_h^2}\Bigg\{\delta(1-z_h) 
\nnu
&
+ \frac{\alpha_s}{2\pi} C_F \bigg[
\frac{2}{\eta} \left(\frac{1}{\epsilon}+\ln\left(\frac{\mu^2}{\mu_b^2}\right)\right) +\frac{1}{\epsilon}\left(\ln\left(\frac{\nu^2}{Q^2}\right)+\frac{3}{2}\right)\bigg]\delta(1-z_h)
\nnu
&+\frac{\alpha_s}{2\pi} \bigg[-\frac{1}{\epsilon}-\ln\left(\frac{\mu^2}{z_h^2\mu_b^2}\right)\bigg]P_{qq}(z_h)
\nnu
&+\frac{\alpha_s}{2\pi} C_F \bigg[\ln\left(\frac{\mu^2}{\mu_b^2}\right) \left(\ln\left(\frac{\nu^2}{Q^2}\right) + \frac{3}{2}\right) \delta(1-z_h) + (1-z_h)\bigg]\Bigg\}\,,
\\[.2cm]
\label{eq:gq}
D_{g/q}(z_h, b, \mu, \nu) =& \frac{1}{z_h^2}\Bigg\{
\frac{\alpha_s}{2\pi} \bigg[-\frac{1}{\epsilon}-\ln\left(\frac{\mu^2}{z_h^2\mu_b^2}\right)\bigg]P_{gq}(z_h)
+\frac{\alpha_s}{2\pi} C_F z_h\Bigg\}\,,
\eea
\end{subequations}
where the splitting functions are given by
\bea
P_{qq}(z_h) = C_F\left[\frac{1+z_h^2}{(1-z_h)_+} + \frac{3}{2}\delta\left(1-z_h\right)\right]\,,
\quad P_{gq}(z_h) = C_F\frac{1+(1-z_h)^2}{z_h}\,.
\eea
The NLO soft function $S_q(b, \mu, \nu)$ can also be computed easily. Since at NLO, only soft radiation that is emitted in the right hemisphere contributes to the hadron transverse momentum $j_T$, this will put a constraint for the soft gluon momentum $k$ in the soft function, i.e., $k_z > 0$ or $k^- > k^+$. 
\bea
{\mathcal S}_q(b, \mu, \nu)=\,&\int d^2{\vec \lambda}_T \, e^{-i{\vec \lambda}_T\cdot {\vec b}}\ \left[\delta^2(\vec{\lambda}_T)+\frac{\alpha_s C_F}{2\pi^2}\frac{ e^{\epsilon\gamma_E}}{\Gamma(1-\epsilon)}\int\frac{ dk^+dk^-}{2}
\right.
\nnu
&\left.\times\left(\frac{\mu^2}{\vec{\lambda}^2_T}\right)^\epsilon\frac{2n\cdot\bar{n}}{k^+k^-}\delta^+(k^+k^--{\vec \lambda_T}^{\,2})\left|\frac{\nu}{2k_z}\right|^\eta\ \theta\left(\frac{k^+}{k^-}>1\right)\right]\nnu
=\,&1 + \frac{\alpha_s}{2\pi} C_F \bigg[ \frac{2}{\eta} \left( - \frac{1}{\epsilon}-\ln\left(\frac{\mu^2}{\mu_b^2}\right)\right) + \frac{1}{\epsilon^2} -\frac{1}{\epsilon} \ln\left(\frac{\nu^2 }{\mu^2}\right)\nnu
& - \ln\left(\frac{\mu^2}{\mu_b^2}\right) \ln\left(\frac{\nu^2 }{\mu_b^2}\right) + \frac{1}{2}\ln^2\left(\frac{\mu^2}{\mu_b^2}\right) - \frac{\pi^2}{12} \bigg]\,.
\label{eq:softb}
\eea
It might be instructive to point out that the above soft function is exactly half of the standard soft function for the back-to-back hadron pair production in $e^+e^-$ collisions, as well as those in SIDIS and Drell-Yan processes. This difference is precisely introduced by the constraint $k_z > 0$ for the radiated soft gluon. This situation is similar to the case where one measures the transverse momentum of hadrons inside a jet with a jet radius $R$, as studied in~\cite{Kang:2017glf}, where soft functions in these two situations are related to each other by a boost along the $z$-direction. 

With the explicit expressions for TMD FFs and soft function at NLO given above, one can easily obtain their corresponding $\mu$ and $\nu$ evolution equations:
\begin{subequations}
\label{eq:evolution}
\bea
\frac{d}{d\ln \mu} \ln D_{h/q}(z_h, b, \mu, \nu) =&\, \gamma_\mu^{D}(\alpha_s)\,,
\\
\frac{d}{d\ln \nu} \ln D_{h/q}(z_h, b, \mu, \nu) =&\, \gamma_\nu^{D}(\alpha_s)\,,
\\
\frac{d}{d\ln \mu} \ln \mathcal{S}_{q}(b, \mu, \nu) =&\, \gamma_\mu^{S}(\alpha_s)\,, 
\\
\frac{d}{d\ln \nu} \ln \mathcal{S}_{q}(b, \mu, \nu) =&\, \gamma_\nu^{S}(\alpha_s)\,.
\eea
\end{subequations}
Here the relevant anomalous dimensions are given by
\bea
\gamma_\mu^{D}(\alpha_s) = & \,\Gamma_{\rm cusp}(\alpha_s) \ln\left(\frac{\nu^2}{Q^2}\right) + 2\gamma^{D_q}(\alpha_s)\,,
\\
\gamma_\mu^{S}(\alpha_s) = & - \Gamma_{\rm cusp}(\alpha_s) \ln\left(\frac{\nu^2}{\mu^2}\right) + \gamma^{S}(\alpha_s)\,,
\\
\gamma_\nu^{D}(\alpha_s) = & - \gamma_\nu^{S}(\alpha_s) = \Gamma_{\rm cusp}(\alpha_s) \ln\left(\frac{\mu^2}{\mu_b^2}\right)\,,
\eea
where the cusp anomalous dimensions $\Gamma_{\rm cusp}$ and the non-cusp $\gamma^{D_q,S}$ have their usual expansion 
\bea
\Gamma_{\rm cusp}(\alpha_s) = \sum_{n=1} \Gamma_{n-1} \left(\frac{\alpha_s}{4\pi}\right)^{n}\,,
\qquad
\gamma^{D_q, S} = \sum_{n=1} \gamma_{n-1}^{D_q,S} \left(\frac{\alpha_s}{4\pi}\right)^{n}\,.
\eea
We have the first few coefficients given by
\bea
& \Gamma_0 =  4C_F\,, 
\qquad 
\Gamma_1 =   \left(\frac{268}{9}-\frac{4 \pi^{2}}{3}\right) C_F C_{A}-\frac{40}{9} C_{F}^2 n_{f}\,, 
\nonumber \\
& \gamma_0^{D_q} =  3\,C_F\,,
\qquad
\gamma_0^{S} =  0\,,
\eea
where $C_F=(N_c^2-1)/(2N_c)$ with $N_c=3$, $C_A = 3$, and $n_f$ represents the quark flavor number.

It is important to realize that the rapidity divergences between TMD FF $D_{h/q}(z_h, b, \mu, \nu)$ and soft function $\mathcal{S}_{q}(b, \mu, \nu)$ cancel between them. This is to be compared with the standard case, e.g., back-to-back hadron pair production in $e^+e^-$ collisions, where the rapidity divergences cancel between one TMD FF $D_{h/q}(z_h, b, \mu, \nu)$ and the square-root of the standard soft function $\sqrt{S_q(b, \mu,\nu)}$, see e.g.~\cite{Collins:2011zzd,Chiu:2012ir,Kang:2015msa,Ebert:2019okf}. Following the modern formulation of TMD FFs, we combine them as the so-called properly-defined TMD FFs~\cite{Collins:2011zzd,Ebert:2019okf} as follows:
\bea
\mathcal{D}^{\rm TMD}_{h/q}(z_h, b, \mu) = D_{h/q}(z_h, b, \mu, \nu) \mathcal{S}_q(b, \mu, \nu)\,.
\eea
Using the evolution equations in Eq.~\eqref{eq:evolution}, one can obtain the evolved TMD FFs $\mathcal{D}^{\rm TMD}$ to be at the hard scale $\mu_h\sim Q$ and thus resum the relevant logarithms $\sim \ln(Q^2/j_T^2)$. For example, the standard exercise is to evolve  $\mathcal{S}_q$ from its characteristic scales $\mu_s \sim \mu_b$ and $\nu_s \sim \mu_b$, and $D_{h/q}$ from its natural scales $\mu_D \sim \mu_b$ and $\nu_D \sim Q$, to the hard scale $\mu_h \sim Q$ and a common rapidity scale $\nu$, from which one obtains
\bea
D_{h/q}(z_h, b, \mu_h, \nu) =\,& D_{h/q}(z_h, b, \mu_b, \nu_D) \left(\frac{\nu}{\nu_D}\right)^{- K(b, \mu_b)} 
\nonumber\\
&\times \exp\left\{\int_{\mu_b}^{\mu_h} \frac{d\mu}{\mu}\left[ \Gamma_{\rm cusp}(\alpha_s)\ln\left(\frac{\nu^2}{Q^2}\right) + 2\gamma^{D_q}(\alpha_s)\right]\right\}\,,
\\
\mathcal{S}_q(b, \mu_h, \nu) = \,&\mathcal{S}_q(b, \mu_b, \nu_s) \left(\frac{\nu}{\nu_s}\right)^{K(b, \mu_b)} 
\nonumber \\
&\times \exp\left\{\int_{\mu_b}^{\mu_h} \frac{d\mu}{\mu}\left[ - \Gamma_{\rm cusp}(\alpha_s)\ln\left(\frac{\nu^2}{\mu^2}\right) + \gamma^{S}(\alpha_s) \right]\right\}\,,
\label{eq:global-soft}
\eea
where $K(b,\mu_b)=\gamma_{\nu}^{S}(\alpha_s)|_{\mu=\mu_b}$ is the rapidity anomalous dimension~\cite{Chiu:2011qc,Chiu:2012ir} or Collins-Soper kernel~\cite{Collins:2011zzd,Ebert:2019okf}. Combine the above evolution equations, we thus obtain 
\bea
\mathcal{D}^{\rm TMD}_{h/q}(z_h, b, \mu_h) = \mathcal{D}^{\rm TMD}_{h/q}(z_h, b, \mu_b)\, e^{-S_{\rm pert}(\mu_b, \mu_h)} \left(\frac{\nu_D}{\nu_s}\right)^{K(b, \mu_b)}\,,
\label{eq:tmd-pert}
\eea
where we have 
\bea
\mathcal{D}^{\rm TMD}_{h/q}(z_h, b, \mu_h) =\,& D_{h/q}(z_h, b, \mu_h, \nu) \mathcal{S}_q(b, \mu_h, \nu)\,,
\\
\mathcal{D}^{\rm TMD}_{h/q}(z_h, b, \mu_b) =\, & D_{h/q}(z_h, b, \mu_b, \nu_D) \mathcal{S}_q(b, \mu_b, \nu_s)\,.
\eea
On the other hand, the exponent of the evolution factor, i.e. the perturbative Sudakov factor $S_{\rm pert}(\mu_b, \mu_h)$ resums all the global logarithms and is given by
\bea
S_{\rm pert}(\mu_b,\mu_h) = \int_{\mu_b}^{\mu_h} \frac{d\mu}{\mu} \left[ \Gamma_{\rm cusp}(\alpha_s) \ln\left(\frac{Q^2}{\mu^2}\right) - 2\gamma^{D_q}(\alpha_s) -\gamma^S(\alpha_s) \right]\,,
\eea
Finally when the scale $\mu_b\gg \Lambda_{\rm QCD}$, one can further match the TMD FFs $\mathcal{D}^{\rm TMD}_{h/q}(z_h, b, \mu_b)$ onto the collinear FFs $D_{h/q}(z_h, \mu_b)$:
\bea
\mathcal{D}^{\rm TMD}_{h/q}(z_h, b, \mu_b)  = \frac{1}{z_h^2} \sum_{i} \int_{z_h}^1 \frac{d z}{z} C_{i \leftarrow q}(z, b,\mu_b) D_{h/i} (z_h/z,\mu_b) + \mathcal{O}\left(\frac{\Lambda_{\rm QCD}^2}{\mu_b^2}\right)\,,
\label{eq:tmd-mub}
\eea
where $C_{i\leftarrow q}(z, \mu_b) = \delta_{iq}\,\delta(1-z)$ at LO and the higher-order expressions can be found in~\cite{Collins:2011zzd,Kang:2015msa,Echevarria:2016scs,Luo:2019hmp}. 

On the other hand, when $\mu_b\sim \Lambda_{\rm QCD}$, one has to introduce non-perturbative contributions, for which we apply the usual $b_{*}$-prescription to include the TMD evolution in the large $b$ region. Here we have $b_{*}$ defined as
\bea
b_{*}=\frac{b}{\sqrt{1+b^2/b_{\rm max}^2}}\,,
\eea
where $b_{\rm max}$ is chosen~\cite{Kang:2015msa} to be 1.5~GeV$^{-1}$. At the same time we include non-perturbative function $S_{\rm NP}(b, Q_0, Q)$, which is given by~\cite{su:2014wpa,Kang:2017glf}
\begin{align}
    S_{\rm NP}(b, Q_0, Q) = \frac{g_2}{2}\ln\left(\frac{b}{b_*}\right)\ln\left(\frac{Q}{Q_0}\right) + \frac{g_h}{z_h^2} b^2\,,
    \label{eq:s-np}
\end{align}
with $Q_0^2=2.4\, {\rm GeV}^2$, $g_2=0.84$ and $g_h=0.042$. We choose to work at the next-to-leading logarithmic (NLL) level, we thus include two-loop cusp and one-loop normal anomalous dimension, and tree-level matching coefficients. Then plugging in the above results for $\mathcal{D}^{\rm TMD}_{h/q}(z_h, b, \mu_h)$ in Eqs.~\eqref{eq:tmd-pert} and \eqref{eq:tmd-mub}, along with the non-perturbative function $S_{\rm NP}(b, Q_0, Q)$ in Eq.~\eqref{eq:s-np}, into the differential cross section in Eq.~\eqref{eq:global-b}, we obtain the all-order resummation formula
\begin{align}
    \frac{d\sigma}{d z_h d^2\vec j_T} &= \sigma_0 \sum_{i=q,\bar q}e_i^2\int_0^\infty \frac{b\, db}{2\pi} J_0(b j_T/z_h) e^{ - S_{\rm pert}(\mu_{b*},\mu_h)-S_{\rm NP}(b, Q_0, Q)}\frac{1}{z_h^2} D_{ h/i}(z_h,\mu_{b*})\,,
    \label{eq:resum-global}
\end{align}
where the Bessel function $J_0$ arises after integrating the angle between $\vec b$ and $\vec j_T$. We have chosen the following scales 
\bea
\mu_h=Q, \qquad 
\mu_{b*}=2e^{-\gamma_E}/b_*\,. 
\eea
Such a formalism in Eq.~\eqref{eq:resum-global} resums all the global logarithms in $\ln(Q^2/j_T^2)$.

\subsection{TMD formalism at threshold $z_h\to 1$}
Belle collaboration finds that the hadron cross sections can be well described by Gaussians in $j_T$ in the small $j_T$ region, and that the width of the Gaussians shows an initially rising for small to intermediate $z_h$, while a decreasing $z_h$-dependence for large $z_h\lesssim 1$. In the region $z_h\to 1$ region, the threshold logarithm of $\ln(1-z_h)$ would become important and thus has to be resummed. In our phenomenological section, we find that the joint threshold and TMD resummation will be able to describe well such a $z_h$-dependence for the Gaussian width. We develop theoretical formalism in this section for this purpose. As we will show below, in the threshold region, an additional mode, so-called collinear-soft (c-soft) mode~\cite{Bauer:2011uc,Procura:2014cba,Lustermans:2016nvk} is relevant. Such a mode is shown as the purple curves in Fig.~\ref{fig:eft} (right), and the corresponding momentum scaling is given by
\begin{itemize}
 \item \makebox[2cm]{{\color{purple} \textbf{c-soft}}:\hfill} $p_{\cSoft} \sim \left( j_T^2/(Q(1-z)), Q(1-z), j_T \right)$ 
\end{itemize}
Let us start our discussion with the fixed-order result of the perturbative TMD FFs $D_{q/q}$ and $D_{g/q}$ in Eq.~\eqref{eq:TMDb} in the threshold limit. By taking the limit $z_h\to 1$, we find at NLO
\bea
D_{q/q}(z_h, b, \mu, \nu) =& \frac{1}{z_h^2}\Bigg\{\delta(1-z_h) 
\nnu
&
+ \frac{\alpha_s}{2\pi} C_F \bigg[
\frac{2}{\eta} \left(\frac{1}{\epsilon}+\ln\left(\frac{\mu^2}{\mu_b^2}\right)\right) +\frac{1}{\epsilon}\left(\ln\left(\frac{\nu^2}{Q^2}\right)+\frac{3}{2}\right)\bigg]\delta(1-z_h)
\nnu
&+\frac{\alpha_s}{2\pi} C_F \bigg[-\frac{1}{\epsilon}-\ln\left(\frac{\mu^2}{\mu_b^2}\right)\bigg]\left[\frac{2}{(1-z_h)_+} + \frac{3}{2}\delta\left(1-z_h\right)\right]
\nnu
&+\frac{\alpha_s}{2\pi} C_F \bigg[\ln\left(\frac{\mu^2}{\mu_b^2}\right) \left(\ln\left(\frac{\nu^2}{Q^2}\right) + \frac{3}{2}\right) \delta(1-z_h) \bigg]\Bigg\}\,,
\eea
where we keep the overall factor of $1/z_h^2$ as a convention. Note that in this limit, one can drop the mixing term $D_{g/q}$ in comparison with the more singular terms $\delta(1-z_h)$ and $\frac{1}{(1-z_h)_+}$ in $D_{q/q}$, i.e. only the flavor diagonal $q\to q$ channel contributes. In the threshold limit, we can refactorize the TMD FF $D_{h/q}$ as
\bea
D_{h/q}(z_h, b, \mu, \nu) = \int_{z_h}^1 \frac{dz}{z} \cSoft_{q}(z, b, \mu, \nu) D_{h/q}(z_h/z, \mu)\,,
\label{eq:D-refac}
\eea
where $\cSoft_{q}$ is a collinear-soft (c-soft) function~\cite{Bauer:2011uc,Procura:2014cba,Lustermans:2016nvk} that takes into account the soft radiation along the direction of the thrust axis, i.e. the c-soft mode mentioned above. At NLO, it can be computed as follows
\bea
\label{eq:c-soft-b}
\cSoft^q(z, b, \mu, \nu)=\,&\delta(1-z)+\frac{\alpha_sC_F}{2\pi^2}\frac{e^{\epsilon\gamma_E}}{\Gamma(1-\epsilon)}\int\frac{dk^+dk^-}{2}\int{d^2\vec{k}_T}e^{-i\vec{k}_T\cdot\vec{b}}\frac{1}{\mu^2}
\nnu
&\times\left(\frac{\mu^2}{\vec{k}_T^2}\right)^{1+\epsilon}\frac{2n\cdot \bar{n}}{k^+k^-}\delta^+(k^2)\delta\left(k^{-} -(1-z)Q\right)\left|\frac{\nu}{2k_z}\right|^\eta 
\nnu
=\,& \delta(1-z) + \frac{\alpha_s}{2\pi}C_F\left[\frac{1}{\epsilon} + \ln\left(\frac{\mu^2}{\mu_b^2}\right)\right]
\left[\left(\frac{2}{\eta}  + \ln\left(\frac{\nu^2}{Q^2}\right)\right) \delta(1-z) - \frac{2}{(1-z)_+}\right]\,.
\eea
Note that the c-soft function $\cSoft^q(z,b,\mu,\nu)$ has the same rapidity anomalous dimension as the TMD FF $D_{h/q}(z_h, b, \mu, \nu)$, which is cancelled after combining soft function $\mathcal{S}_q$ in Eq.~\eqref{eq:softb} and the c-soft function in Eq.~\eqref{eq:c-soft-b}. On the other hand, we also have the collinear FFs at the threshold limit, whose perturbative expression is given by
\bea
D_{q/q}(z_h, \mu) = \delta(1-z_h) + \frac{\alpha_s}{2\pi}C_F\left(-\frac{1}{\epsilon}\right)\left[\frac{2}{(1-z_h)_+} + \frac{3}{2}\delta\left(1-z_h\right)\right]\,.
\eea

To perform the resummation in the threshold limit, one can perform the Mellin transform or Laplace transformation~\cite{Sterman:2013nya}, whose purpose is to convert the above convolution in $z$-space into a simple product in the corresponding transformed space. Here we choose to perform the Laplace transformation~\cite{Becher:2006nr}, 
\bea
\tilde D_{h/q}(\kappa, b, \mu, \nu) = \int_0^\infty d\bar z_h e^{-\frac{\bar z_h}{\kappa e^{\gamma_E}}} D_{h/q}(1-\bar z_h, b, \mu, \nu)\,,
\eea
where $\bar z_h = 1 - z_h$. Using the following relation in the threshold limit
\bea
1-z_h = 1 - \left[1 - \left(1-\frac{z_h}{z}\right)\right] \left[1 - \left(1-z\right)\right] \approx \left(1-\frac{z_h}{z}\right) + \left(1-z\right)\,,
 \eea
one can express Eq.~\eqref{eq:D-refac} as a product in the Laplace space 
\bea
\tilde D_{h/q}(\kappa, b, \mu, \nu) = \tilde \cSoft_{q}(\kappa, b, \mu, \nu) \tilde D_{h/q}(\kappa, \mu)\,.
\eea
Note that we have also extended the integration from $0 < \bar z < 1$ to $0 < \bar z < \infty$ in the threshold limit approximation. The NLO expressions for $\tilde \cSoft_{q}(\kappa, b, \mu, \nu)$ and $\tilde D_{h/q}(\kappa, \mu)$ in the Laplace space are given by
\bea
\tilde \cSoft^q(\kappa, b, \mu, \nu) =\,& 1+ \frac{\alpha_s}{2\pi}C_F\left[\frac{1}{\epsilon} + \ln\left(\frac{\mu^2}{\mu_b^2}\right)\right]
\left[\frac{2}{\eta} + \ln\left(\frac{\nu^2}{\kappa^2 Q^2}\right)\right]\,,
\\
\tilde D_{q/q}(\kappa, \mu) = \, & 1 + \frac{\alpha_s}{2\pi}C_F \left(-\frac{1}{\epsilon}\right) \left[\ln\left(\kappa^2\right) + \frac{3}{2}\right]\,.
\eea 
From the above results, one can derive the RG equations for both $\tilde \cSoft^q$ and $D_{q/q}$ in the Laplace space, 
\begin{align}
    & \frac{d}{d\ln\mu} \ln \tilde{\cSoft}_q(\kappa, b, \mu, \nu) = \left[ \Gamma_{\rm cusp}(\alpha_s) \ln\left(\frac{\nu^2}{\kappa^2 Q^2}\right) + \gamma^{\tilde{\cSoft}_q}(\alpha_s)\right]\,, \\
        &\frac{d}{d\ln \nu} \ln \tilde{\cSoft}_q(\kappa, b, \mu, \nu) = \gamma^D_\nu(\alpha_s)\,, \\
    &\frac{d}{d\ln\mu} \ln \tilde D_{h/q}\left(\kappa,\mu\right) = \left[ \Gamma_{\rm cusp}(\alpha_s) \ln\left(\kappa^2\right) + 2\gamma^{f_q}(\alpha_s)\right]  \,,
\end{align}
where the normal anomalous dimensions $\gamma^{i}$ expanded as $\gamma^{i} = \sum_{n=1}\gamma_{n-1}^{i}\left(\alpha_s/4\pi\right)^n$ with $i=\tilde\cSoft_q, f_q$, and
\bea
\gamma_0^{f_q}=  3\,C_F\,, \qquad
\gamma_0^{\tilde\cSoft_q} =  0\,.
\eea

The above RG equations allow us to evolve c-soft function $\tilde\cSoft^q(\kappa, b, \mu, \nu)$ from its natural scale $\mu_\cSoft \sim \mu_b$ and $\nu_\cSoft\sim \kappa Q$, and the FF $D_{h/q}(\kappa, \mu)$ from initial scale $\mu_F$, up to the hard scale $\mu_h$ and a rapidity scale $\nu$, we obtain
\bea
\tilde \cSoft^q(\kappa, b, \mu_h, \nu) =\,& \tilde \cSoft^q(\kappa, b, \mu_b, \nu_\cSoft)  \left(\frac{\nu}{\nu_\cSoft}\right)^{- K(b, \mu_b)}
\nonumber \\
&\times 
\exp \left\{\int_{\mu_b}^{\mu_h} \frac{d\mu}{\mu} \left[\Gamma_{\rm cusp}(\alpha_s) \ln\left(\frac{\nu^2}{\kappa^2 Q^2}\right) + \gamma^{\tilde{\cSoft}_q}(\alpha_s)\right]\right\}\,,
\\
D_{h/q}(\kappa, \mu_h) = \, & D_{h/q}(\kappa, \mu_F) \exp\left[ \int_{\mu_F}^{\mu_h} \frac{d\mu}{\mu} \left(\Gamma_{\rm cusp}(\alpha_s)\ln(\kappa^2)+2 \gamma^{f_q}(\alpha_s) \right)\right]\,.
\eea
Combining the evolution for the global soft function $S_{q}(b, \mu, \nu)$ in Eq.~\eqref{eq:global-soft}, we obtain the following evolution for the properly-defined TMD FFs in the Laplace space, 
\bea
\tilde {\mathcal D}_{h/q}^{\rm TMD}(\kappa, b, \mu_h) = \, & S_q(b, \mu_b, \nu_s) \tilde \cSoft^q(\kappa, b, \mu_b, \nu_\cSoft) \tilde D_{h/q}(\kappa, \mu_F) 
\nonumber\\
\,&
\times e^{-\tilde S_{\rm pert}(\mu_b, \mu_h)}  
\left(\frac{\nu_\cSoft}{\nu_s}\right)^{K(b, \mu_b)}\,.
\eea
Here the perturbative Sudakov factor $\tilde S_{\rm pert}(\mu_b, \mu_h)$ is given by
\begin{align}
\label{r2suda}
     \tilde S_{\rm pert}(\mu_b,\mu_h) =\,&  \int_{\mu_b}^{\mu_h} \frac{d\mu}{\mu} \left[ \Gamma_{\rm cusp}(\alpha_s) \ln\left(\frac{\kappa^2 Q^2}{\mu^2}\right)  \right]
     \nonumber \\
     \, & - \int_{\mu_F}^{\mu_h} \frac{d\mu}{\mu} \left[ \Gamma_{\rm cusp}(\alpha_s) \ln\left(\kappa^2\right) + 2\gamma^{f_q}(\alpha_s) \right]\,,
\end{align}
where the first integral represents the evolution of c-soft function and the global part of the soft function from $\mu_b$ to $\mu_h$, and the second one is collinear fragmentation function from factorization scale $\mu_F$ to $\mu_h$ in the threshold limit. Performing the inverse Laplace transform, we obtain the following expression for TMD FFs in the threshold limit
\bea
{\mathcal D}_{h/q}^{\rm TMD}(z_h, b, \mu_h) = \frac{1}{z_h^2} \int_{z_h}^1\frac{dz}{z} e^{- \hat S_{\rm pert}(\mu_b, \mu_h)}  
\frac{e^{-2\gamma_E \eta}}{\Gamma(2\eta)} \frac{1}{1-z} D_{h/q}(z_h/z, \mu_h)\,.
\eea
Here we derive the above formula using the first line of Sudakov factor in Eq.~\eqref{r2suda} and setting $\mu_F=\mu_h$, and the parameter $\eta$ is defined as
\bea
\eta= - \int_{\mu_{b}}^{\mu_h} \frac{d\mu}{\mu} \Gamma_{\rm cusp}(\alpha_s)\,.
\eea
On the other hand, $\hat S_{\rm pert}(\mu_b, \mu_h)$ in the momentum space in the threshold limit is given by
\bea
\hat S_{\rm pert}(\mu_b, \mu_h) = \int_{\mu_b}^{\mu_h} \frac{d\mu}{\mu} \left[\Gamma_{\rm cusp}(\alpha_s) \ln\left(\frac{(1-z)^2Q^2}{\mu^2}\right)\right]\,,
\label{eq:threshold-spert}
\eea
where the argument in the logarithm is given by $(1-z) Q$ instead of the usual $Q$ in the threshold limit. 

Finally using the above result, one can obtain the resummed formalism for the differential cross section at the NLL level
\begin{align}
\label{jointres}
    \frac{d\sigma}{d z_h d^2\vec j_T} =\,& \sigma_0
    \sum_{i=q,\bar q}\int_0^\infty \frac{b\, db}{2\pi} J_0(b j_T/z_h) 
    \notag \\
    &\,\times \frac{1}{z_h^2} \int_{z_h}^1 \frac{dz}{z} e^{ - \hat S_{\rm pert}(\mu_{b*},\mu_h) - \hat S_{\rm NP}(b, Q_0, Q)}  \frac{e^{-2\gamma_E \eta}}{\Gamma(2\eta)} \frac{1}{1-z} D_{h/i}(z_h/z,\mu_h)\,,
\end{align}
where we choose the non-perturbative Sudakov factor $\hat S_{\rm NP}(b, Q_0, Q)$ to have the following form
\begin{align}
   \hat S_{\rm NP}(b, Q_0, Q) = \frac{g_2}{2}\ln\left(\frac{b}{b_*}\right)\ln\left[\frac{Q(1-z_h)}{Q_0}\right] + \frac{g_h}{z_h^2} b^2\,.
\end{align}
Here motivated by the argument in the perturbative Sudakov function in Eq.~\eqref{eq:threshold-spert}, we replace $Q$ by $(1-z_h)Q$ in the usual non-perturbative function $S_{\rm NP}(b, Q_0, Q)$ in Eq.~\eqref{eq:s-np} to obtain $\hat S_{\rm NP}(b, Q_0, Q)$ in the threshold limit. 

\section{Factorization and Resummation: full story}
\label{sec:full}

The plane perpendicular to the thrust axis divides the full space into two hemispheres. One measures the transverse momentum of hadron $h$ in the right hemisphere, and the left hemisphere is inclusive. As we have emphasized above, hadron transverse momentum with respect to the thrust axis is a non-global observable, since the left hemisphere are unobserved. Such type of observables will involve non-global structures which can not be captured by the traditional exponential formula~\cite{Sterman:2004en}. In this section, we apply formalism developed in~\cite{Becher:2016omr,Becher:2016mmh} for the jet effective theory to derive the factorization and resummation formula. Such a formalism will enable us to resum both global and non-global logarithms. The global logarithmic structure has been discussed in the previous section. Here in this section, we pay more attention to the NGLs~\cite{Dasgupta:2002bw}. Very recently, a similar structure is also mentioned in \cite{1808927}.

In the standard TMD region where $j_T \ll Q$, following the development in~\cite{Becher:2016omr,Becher:2016mmh}, we can write the factorization formalism as follows 
\begin{align}\label{r1fac_mom}
    &\frac{d\sigma}{d z_h d^2 \vec j_T } = \sum_{i=q,\bar q, g} \int d^{2} \vec{k}_T d^{2} \vec{\lambda}_T \, \delta^{(2)}(\vec{j}_T-\vec k_T- z_h \vec \lambda_T) \notag \\ 
    &\hspace{1cm}\times\sum_{m=2}^\infty \frac{1}{N_c}{\rm Tr}_c\Big[  \bm{\mathcal{H}}_m^i(\{\underline{n}\},Q,\mu)\otimes \bm{\mathcal{S}}_m(\{\underline{n}\},\lambda_T,\mu,\nu) \Big] D_{h/i}(z_h,k_T,\mu,\nu)\,,
\end{align}
where $\bm{\mathcal{H}}$, $\bm{\mathcal{S}}$, and $D_{h/i}$ correspond to hard, soft and TMD FFs, respectively. Besides, different from the formalism in the previous section that resums only global logarithms, the hard and soft functions are now matrices in the color space, so we take color averaging as ${\rm Tr}_c[\cdots]/N_c$ after multiplying them and integrating out the solid angles $\{\underline{n}\}=\{n_1,n_2,\cdots\}$ of the hard partons, where the angular integration is expressed by the symbol $\otimes$. The index $m$ denotes the number of energetic partons inside the hard function that is defined in \cite{Becher:2017nof}. The index $m$ in soft function then represents the number of Wilson lines, and the momentum space soft function is defined as 
\begin{align}
\bm{\mathcal{S}}_{m}(\{\underline{n}\}, \lambda_{T})=\,& \int \!\!\!\!\!\!\!\!\!\sum_{X_{s}}  \delta^{(2)}\left(\vec{p}_{X_{R}}^{\perp}-\vec \lambda_{T}\right) \nonumber \\ 
&\hspace{-8mm} 
\times\langle 0|\boldsymbol{S}_{0}^{\dagger}(n) \boldsymbol{S}_{1}^{\dagger}\left(n_{1}\right) \ldots \boldsymbol{S}_{m}^{\dagger}\left(n_{m}\right)| X_{R}\rangle\left\langle X_{R}\left|\boldsymbol{S}_{0}(n) \boldsymbol{S}_{1}\left(n_{1}\right) \ldots \boldsymbol{S}_{m}\left(n_{m}\right)\right| 0\right\rangle. 
\end{align}
Here $X_{R}$ denotes soft states in the right hemisphere, and one only measures the contributions from soft radiations in the right hemisphere. It is precisely because of such a multi-Wilson line structure that makes the hard and soft function matrices in the color space. After performing Fourier transformation for the observed $\vec j_T$, the factorization formula is given as
\begin{align}
\label{r1fac}
    \frac{d\sigma}{d z_h d^2 \vec j_T } =\,& \sum_{i=q,\bar q, g} \int \frac{d^2 \vec b}{(2\pi)^2} e^{i \vec b\cdot \vec j_T/z_h} 
    \notag \\ 
   &\, \times\sum_{m=2}^\infty \frac{1}{N_c}{\rm Tr}_c\Big[  \bm{\mathcal{H}}_m^i(\{\underline{n}\},Q,\mu)\otimes \bm{\mathcal{S}}_m(\{\underline{n}\},b,\mu,\nu) \Big] D_{h/i}(z_h,b,\mu,\nu)\,.
\end{align}

For the NGLs resummation, we use the same methods in \cite{Becher:2016omr} to perform renormalization for the multi-Wilson-line operators. The renormalization constants of the hard and soft function are matrices in the color space, which are given as 
\begin{align}
\bm{\mathcal{H}}_{m}\left(\{\underline{n}\}, Q, \epsilon\right)=\,&\sum_{l=2}^{m} \bm{\mathcal{H}}_{l}\left(\{\underline{n}\}, Q, \mu\right) \boldsymbol{Z}_{l m}^{H}(\{\underline{n}\}, \mu, \epsilon)\,,
\\
    \bm{\mathcal{S}}_{l}\left(\{\underline{n}\}, b, \mu,\nu\right)=\,&\sum_{m=l}^{\infty} \bm{Z}_{l m}^{S}\left(\{\underline{n}\}, b, \mu,\nu, \epsilon,\eta\right) \hat{\otimes} \,\bm{\mathcal{S}}_{m}\left(\{\underline{n}\}, b, \epsilon,\eta\right)\,,
\end{align}
separately. The symbol $\hat \otimes$ denotes the integration of the angular vectors $\{n_{l+1},n_{l+2},\cdots\}$ defined in \cite{Becher:2016mmh}. The factor $\bm{Z}^H$ and $\bm{Z}^S$ are connected through the renormalization factor ($Z^D$) of the TMD FF as $ \bm{Z}^S = Z^D \bm{Z}^H $. The above relations have been verified at the two-loop order in \cite{Becher:2016omr, Becher:2017nof}. For convenience we define the global renormalization constant $Z^S$ for the soft function as \cite{Chien:2019gyf} 
\begin{align}
    \bm{\mathcal{S}}_{2}\left(\{\underline{n}\}, b, \epsilon,\eta \right) = \bm{\mathcal{S}}_{2}\left(\{\underline{n}\}, b, \mu,\nu\right) Z^S.
\end{align}
Then the global hard renormalization constant $Z^H$ is given as $Z^H=(Z^D Z^S)^{-1}$, and the non-global renormalization constant $\bm{\hat Z}_{lm}$ is given as
\begin{align}
    \bm{\hat Z}_{lm} = \bm{Z}_{lm}^S Z^S,~~~~ \bm{Z}_{lm}^H=\bm{\hat Z}_{lm} Z^H.
\end{align}
Based on the above definitions, the RG equations of each ingredients are given as
\begin{align}
    &\frac{d}{d\ln\mu} \bm{\mathcal{H}}_{m}\left(\{\underline{n}\}, Q, \mu\right) =  \\
    & \hspace{0.5cm} \sum_{l=2}^m \bm{\mathcal{H}}_{l}\left(\{\underline{n}\}, Q, \mu\right) \left\{ \left[\Gamma_{\rm cusp}(\alpha_s) \ln \frac{Q^2}{\mu^2} - 2\gamma^{D_q}(\alpha_s) - \gamma^S(\alpha_s)\right] \delta_{lm} \bm{1} - \bm{\hat \Gamma}_{lm} (\{\underline{n}\},\mu)\right\}, \notag\\
    &\frac{d}{d\ln\mu} \bm{\mathcal{S}}_{l}\left(\{\underline{n}\}, b, \mu,\nu\right) =  \\
    & \hspace{1.1cm}\sum_{m=l}^\infty \left\{ \left[ -\Gamma_{\rm cusp}(\alpha_s) \ln\frac{\nu^2}{\mu^2} +\gamma^S(\alpha_s)\right] \delta_{lm} \bm{1} + \bm{\hat \Gamma}_{lm} (\{\underline{n}\},\mu)\right\} \hat \otimes\, \bm{\mathcal{S}}_{m}\left(\{\underline{n}\}, Q, \mu,\nu\right), \notag\\
    &\frac{d}{d\ln\mu} D_{h/i}(z,b,\mu,\nu) = \left[ \Gamma_{\rm cusp}(\alpha_s) \ln\frac{\nu^2}{Q^2} + 2\gamma^{D_q}(\alpha_s)\right]  D_{h/i}(z,b,\mu,\nu).
\end{align}
where the anomalous dimensions are derived via
\begin{align}
    \Gamma = - Z^{-1} \frac{d}{d\ln\mu} Z.
\end{align}
Besides, both soft and TMD FF are suffering from the rapidity divergence, and the corresponding Rapidity-RG equations for them are
\begin{align}
    &\frac{d}{d\ln\nu} \bm{\mathcal{S}}_{l}\left(\{\underline{n}\}, b, \mu,\nu\right) =  \gamma^S_\nu(\alpha_s)  \bm{\mathcal{S}}_{l}\left(\{\underline{n}\}, b, \mu,\nu\right). \\
    &\frac{d}{d\ln\nu} D_{h/i}(z,b,\mu,\nu) = \gamma^D_\nu(\alpha_s)  D_{h/i}(z,b,\mu,\nu).
\end{align}
Similarly, the rapidity anomalous dimension is defined as
\begin{align}
    \gamma_\nu = - Z^{-1} \frac{d}{d\ln\nu} Z.
\end{align}
The expressions for the one-loop global anomalous dimensions have been given in the previous section. After solving the RG equations, we can obtain all-order resummation formula. At the NLL accuracy it has the form as
\bea
\label{eq:resum-full}
    \frac{d\sigma}{d z_h d^2\vec j_T} =\,& \sigma_0 \sum_{i=q, \bar q} e_i^2 \int_0^\infty \frac{b\, db}{2\pi} J_0(b j_T/z_h) e^{ - S_{\rm pert}(\mu_{b*},\mu_h)-S_{\rm NP}(b, Q_0, Q)}
    \nonumber \\
   &\times \frac{1}{z_h^2} D_{ h/i}(z_h,\mu_{b*})U_{\rm NG} (\mu_{b*},\mu_h)\,.
\eea
In comparison with the resummed formalism in Eq.~\eqref{eq:resum-global}, we have the non-global evolution function $U_{\rm NG}$, which is given as
\begin{align}
\label{eq:UNG}
    &U_{\rm NG}(\mu_{b*},\mu_h) = \notag \\ &\hspace{1cm}\frac{1}{N_c}\sum_{l=2}^{\infty}{\rm Tr}_c\Big[\bm{\mathcal{H}}_{l}\left(\left\{\underline{n}^{\prime}\right\}, Q, \mu_{h}\right) \otimes \sum_{m \geq l}^{\infty} \boldsymbol{U}_{l m}\left(\{\underline{n}\}, \mu_{h}, \mu_{b*}\right) \hat{\otimes}\, \bm{\mathcal{S}}_{m}\left(\{\underline{n}\}, b, \mu_{b*}\right)\Big ],
\end{align}
where $U_{\rm NG}$ is the evolution function for the non-global parts. At the LL accuracy and the large-$N_c$ limit, one can calculate it using the parton shower algorithms in \cite{Dasgupta:2001sh,Balsiger:2018ezi} or the numerical solution of the BMS equations \cite{Banfi:2002hw}. For the convenience of our numerical calculations in the next section, however, we choose the parametrization given in~\cite{Dasgupta:2001sh}
\begin{align}
    U_{\mathrm{NG}}\left(\mu_{b*}, \mu_{h}\right) = \exp \left[-C_{A} C_{F} \frac{\pi^{2}}{3} u^{2} \frac{1+(a u)^{2}}{1+(b u)^{c}}\right],
\end{align}
with $a=0.85 \, C_{A}$, $b=0.86 \, C_{A}$, $c=1.33$, and 
\begin{align}
    u= \int_{\mu_b*}^{\mu_h} \frac{d\mu}{\mu} \frac{\alpha_s(\mu)}{2\pi} = \frac{1}{\beta_0}\ln \left[ \frac{\alpha_{s}\left(\mu_{b*}\right)}{\alpha_{s}\left(\mu_{h}\right)}\right]\,,
\end{align}
where $\beta_0=\frac{11}{3}C_A-\frac{4}{3}T_Fn_f$, with $T_F = 1/2$.

For the differential cross section in the threshold limit, we find that the same non-global evolution function $U_{\rm NG}$ arises. We thus write the resummed formalism at the NLL in the threshold limit $z_h\to 1$ as
\begin{align}
\label{jointres-full}
    \frac{d\sigma}{d z_h d^2\vec j_T} =\,& \sigma_0
    \sum_{i=q,\bar q} e_i^2 \int_0^\infty \frac{b\, db}{2\pi} J_0(b j_T/z_h) \int_{z_h}^1 \frac{dz}{z} e^{ - \hat S_{\rm pert}(\mu_{b*},\mu_h) - \hat S_{\rm NP}(b, Q_0, Q)}  
    \notag \\
    &\,\times \frac{1}{z_h^2} \frac{e^{-2\gamma_E \eta}}{\Gamma(2\eta)} \frac{1}{1-z} D_{h/i}(z_h/z,\mu_h) U_{\rm NG} (\mu_{b*},\mu_h)\,.
\end{align}

\section{Numerical results}
\label{sec:pheno}

\begin{figure}[t]
\begin{center}
  \includegraphics[height=0.9\textwidth]{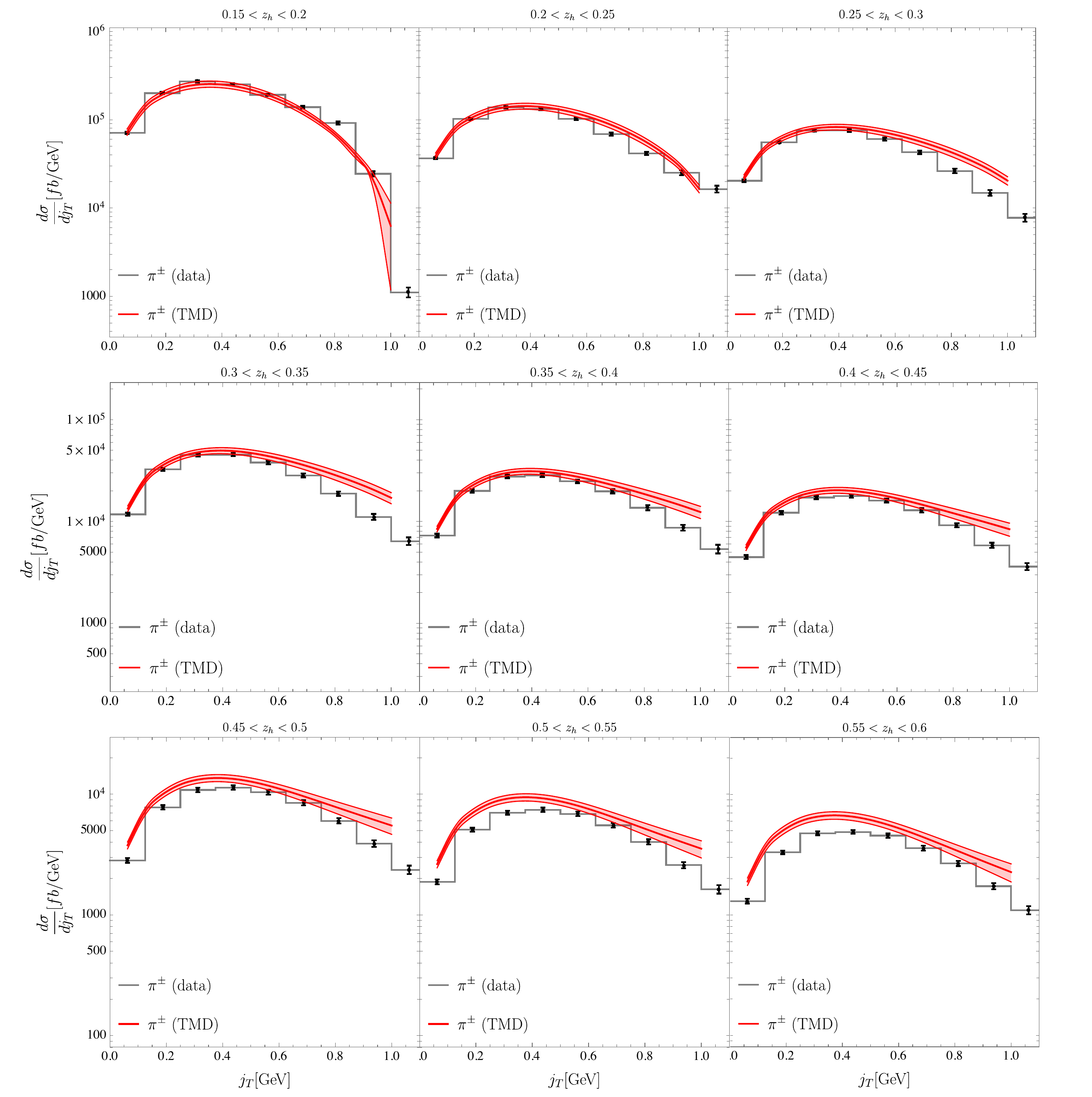}
\end{center}
  \caption{Differential cross sections for the charged pion as a function of $j_T$ and different $z_h$ bins with $\Delta z_h=0.05$ interval in the region $0.15 < z_h < 0.6$. In each plot, TMD resummation is applied and data points (black dots and histogram with error bars) from Belle collaboration \cite{Seidl:2019jei} are shown for comparison. The error bands correspond to 90\% C.L. uncertainty obtained from DSS fit \cite{de_Florian_2015}.}
\label{fig:qTdis}
\end{figure}

In this section, we will study the differential cross sections and Gaussian widths of the transverse momentum distribution for the single inclusive charged pion production (sum of $\pi^+$ and $\pi^-$) in electron-positron annihilation process, $e^+e^-\rightarrow \pi^\pm+X$, based on the factorization and NLL resummation formula given in Sec.~\ref{sec:full}. For parton-to-pion fragmentation functions, we use 2014 DSS analysis \cite{de_Florian_2015}, where the uncertainties were determined based on the standard iterative Hessian method. Note that Belle data~\cite{Seidl:2019jei} was originally presented in different thrust bins, in $0.5<T<0.7$, $0.7<T<0.8$, $0.8<T<0.9$, $0.9<T<0.95$ and $0.95<T<1.0$. Since the theoretical formalism we have developed in this paper is inclusive in the thrust variable, we thus combine the experimental data to obtain the results for the entire region $0.5 < T < 1.0$. The data shown in this section are all the ones obtained via such a combination procedure. The errors of the data sets are also combined weighted by corresponding thrust bins.

Fig.~\ref{fig:qTdis} shows the differential cross sections for pion production in $e^+e^-$ collision as a function of the pion transverse momentum $j_T$, in different $z_h$ bins at the center-of-mass energy $\sqrt{s}=10.58$ GeV. The error bands correspond to 90\% confidence level (C.L.) uncertainty of parton-to-pion FFs determined in~\cite{de_Florian_2015}. The hadron transverse momentum with respected to thrust axis are given in $0<j_T<1.0$ GeV for each plot. The energy fraction region $0.1<z_h<0.65$ is divided into eleven sub-regions with $\Delta z_h=0.05$ for each panel. As seen clearly in the figure, for the intermediate $z_h$ region ($z_h\lesssim 0.5$), the evaluations based on TMD resummation in Eq.~\eqref{eq:resum-full} are in good agreement with the data~\footnote{We have included an overall normalization of 0.25 in our theory to match the experimental data. Such a normalization factor is consistent with what is fitted in~\cite{Soleymaninia:2019jqo}.}. On the other hand, as $z_h$ becomes relatively large ($z_h\gtrsim 0.5$) and thus approaches threshold limit, the agreement becomes worse, which indicates the potential importance of the threshold resummation effect.

\begin{figure}[t]
\begin{center}
  \includegraphics[height=0.3\textwidth]{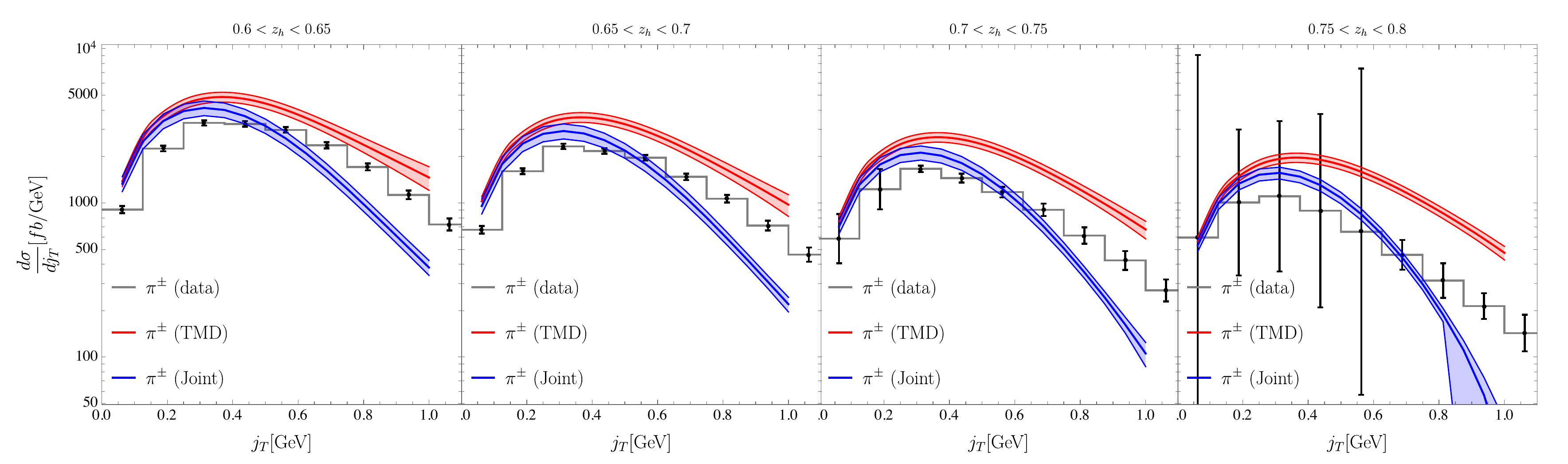}
\end{center}
  \caption{Transverse momentum $j_T$ distribution given by TMD resummation (red band) and the joint TMD and threshold resummation (blue band) for the charged pion production with $z_h$ bins $0.65<z_h<0.7$, $0.7<z_h<0.75$, $0.75<z_h<0.8$ and $0.8<z_h<0.85$ from left to right. Results are shown in comparison with Belle data \cite{Seidl:2019jei} in each pad. Error bands represent 90\% C.L. uncertainty.}
\label{fig:qTjoint}
\end{figure}

In Fig. \ref{fig:qTjoint} we compare the differential cross sections obtained by using two resummations schemes: transverse momentum resummation (shown in red curves) and joint transverse momentum and threshold resummation (shown in blue curves). Similarly as before, the error bands correspond to 90\% confidence level (C.L.) uncertainty of parton-to-pion FFs. The hadron transverse momentum with respected to thrust axis are given in $0<j_T<1.0$ GeV region. The energy fraction regions are $0.65<z_h<0.7$, $0.7<z_h<0.75$, $0.75<z_h<0.8$ and $0.8<z_h<0.85$ from left to right. In Fig. \ref{fig:qTjoint}, $z_h$ bins are larger than those in Fig. \ref{fig:qTdis} where the threshold logarithms are making some difference, thus compared to TMD resummation, we see that joint resummation has a better performance in these $z_h$ bins, especially in the small $j_T$ region. As $z_h$ gets larger, the consistency between joint resummation results and data gets better with a decreasing Gaussian width.  The jointly resummed differential cross section decreases faster, indicating the a smaller Gaussian width value, which is more consistent with experimental data compared to the results with only transverse momentum resummed, where shapes are almost the same for the four $z_h$ bins in such a large $z_h$ region.

\begin{figure}[t]
\begin{center}
  \includegraphics[height=0.45\textwidth]{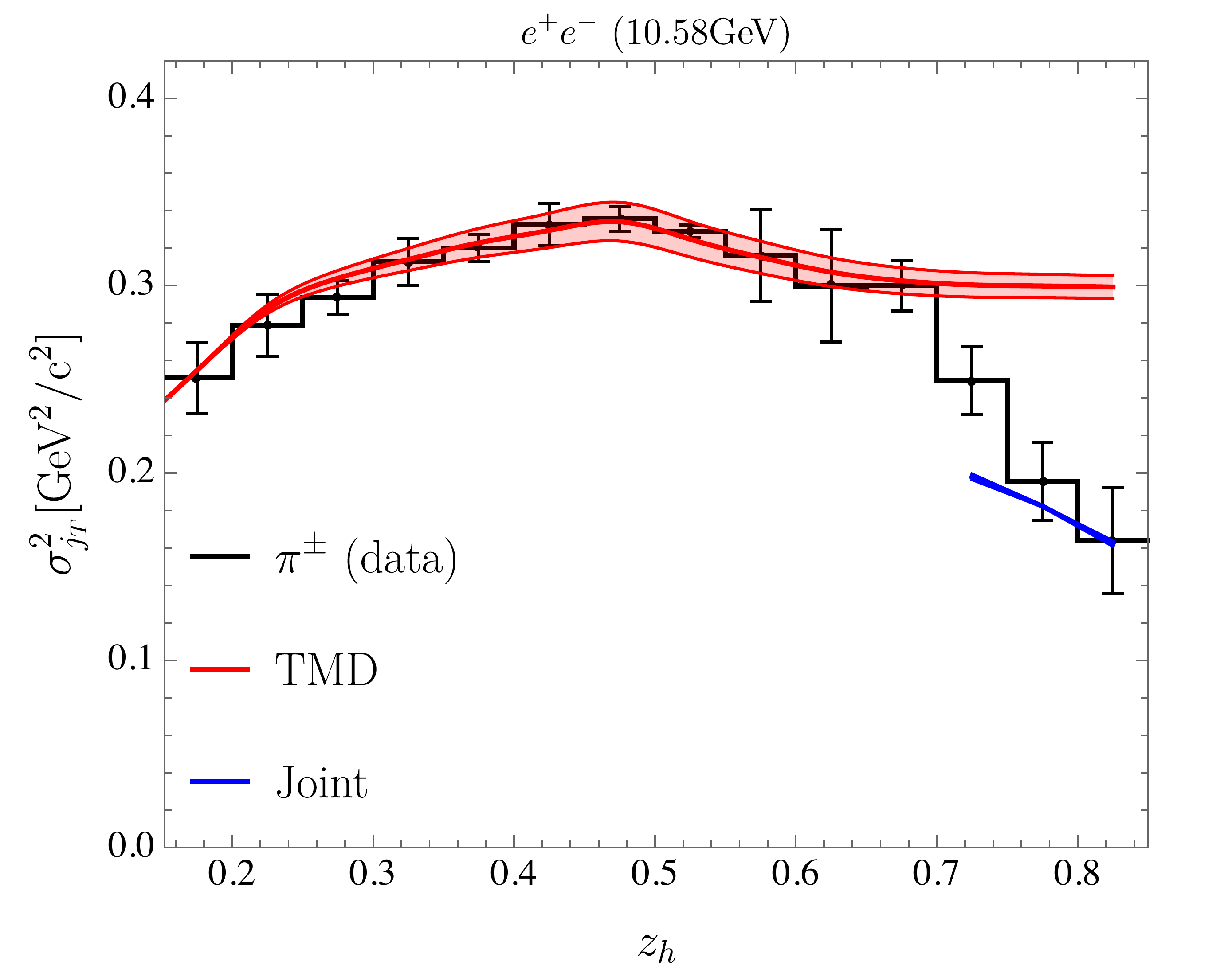}
\end{center}
  \caption{Gaussian widths for pion using TMD resummation (red band) and joint resummation (blue band) as a function of $z_h$ in the thrust bin $0.5 < T < 1.0$ under 90\% C.L. Data points are constructed by charged pion differential cross sections measured at Belle detector. }
\label{fig:gauss}
\end{figure} 

To see the change of $j_T$ width as a function of $z_h$, we fit the cross section $d\sigma/dz_h d^2\vec{j}_T$ as a function of $j_T^2$,
\bea
\frac{d\sigma}{dz_h d^2\vec{j}_T} \propto \frac{1}{\pi \sigma_{j_T}^2} \exp\left(-j_T^2/ \sigma_{j_T}^2\right)\,,
\eea
and reconstruct the Gaussian width $\sigma_{j_T}^2$ for both theory and experimental data. We compute Gaussian width as a function of fractional energy $z_h$ using both TMD resummation (red curve) and joint resummation (blue curve). In Fig.~\ref{fig:gauss}, for small $z_h$ region ($z_h<0.5$), the logarithmically enhanced contribution origins from $\ln(Q/j_T)$, thus transverse momentum resummed cross section $\sigma^2_{j_{T}}$ fits the data well. As the value of $z_h$ is increased, for the TMD factorization theorem in Eq.~\eqref{eq:resum-full}, dependence on $z_h$ becomes weak, leading to a plateau at the tail region. On the other hand, for the factorization theorem with joint resummation in Eq.~\eqref{jointres-full}, where transverse momentum and threshold logarithms are jointly resummed, the cross section sharply decreases as $z_h$ increases, indicating a better fit for this region.
Generally speaking, for kinematic regions distinguished by $z_h$ bins, adopting TMD resummation in intermediate $z_h$ regions while making use of joint resummation for large $z_h$ bins can lead to excellent agreement with measurement for $e^+e^-\rightarrow \pi X$ process, suggesting our factorization and resummation formula results in a reasonable approach for describing single inclusive hadron production at the electron-positron colliders.

\section{Conclusion}
\label{sec:summary}

Single inclusive hadron production at the $e^+e^-$ colliders provide a new opportunity to study transverse momentum dependent fragmentation functions (TMD FFs), which are important to understand the 3D structure for the hadrons and the non-perturbative QCD. Belle collaboration has performed the first measurement for this observable, $e^+e^-\to h(z_h, j_T) + X$, where $z_h$ is the energy fraction for the hadron, while the hadron's transverse momentum $j_T$ is measured with respect to the thrust axis determined by the hadronic event shape. We develop a TMD factorization formalism for such an observable, which resums logarithm of $\ln(Q/j_T)$. Realizing the non-global nature of the observable, our factorization formalism involves a multi-Wilson line structure, which allows us to resum both global and non-global logarithms. Besides, as the increasing of the energy fraction $z_h$ of the hadron, the threshold soft gluon enhancement effects become more and more important, which require us to perform joint TMD ($\sim \ln(Q/j_T)$) and threshold ($\sim\ln(1-z_h)$) resummation. We apply the formalism proposed in \cite{Lustermans:2016nvk} based on SCET+ framework \cite{Bauer:2011uc} to obtain factorization and resummation formula in the joint limit.

 In the end we find that TMD resummation formula give a good description for the $j_T$ distribution as $z_h<0.65$. For large $z_h>0.65$ region, in order to describe the data we need to include threshold resummation effects. Especially, we find that the Gaussian width of the $j_T$ distribution given by the TMD formalism freeze to a certain value which is not consistent with the measurement. While after including joint threshold and TMD resummation effects, the theoretical predictions are consistent with data very well.

In the present work we obtained the perturbative resummed cross section at the next-to-leading logarithmic (NLL) accuracy. In the future work, we will include higher order resummation effects using method developed in~\cite{Balsiger:2019tne}. Especially, in this case beyond the NLL level, the gluon TMD FF will also contribute to the cross section as shown in \eqref{r1fac}, it will be interesting to study its effects. 

\acknowledgments

We thank Ralf Seidl for discussions on the combination of Bell data with different thrust bins, and thank Daniele Anderle and Anselm Vossen for useful discussions. This work is supported by the National Science Foundation under Grant No.~PHY-1720486 and CAREER award~PHY-1945471 (Z.K., D.Y.S., and F.Z.), and by Center for Frontiers in Nuclear Science of Stony Brook University and Brookhaven National Laboratory (D.Y.S.). This work is also supported within the framework of the TMD Topical Collaboration.

\bibliographystyle{JHEP}
\bibliography{jet}

\providecommand{\href}[2]{#2}\begingroup\raggedright\begin{thebibliography}{10}

\bibitem{Accardi:2012qut}
A.~Accardi et~al., {\it {Electron Ion Collider: The Next QCD Frontier}:
  {Understanding the glue that binds us all}},  {\em Eur. Phys. J. A} {\bf 52}
  (2016), no.~9 268, [\href{http://arxiv.org/abs/1212.1701}{{\tt
  arXiv:1212.1701}}].

\bibitem{Boer:2011fh}
D.~Boer et~al., {\it {Gluons and the quark sea at high energies: Distributions,
  polarization, tomography}},  \href{http://arxiv.org/abs/1108.1713}{{\tt
  arXiv:1108.1713}}.

\bibitem{Aidala:2020mzt}
C.~A. Aidala et~al., {\em {Probing Nucleons and Nuclei in High Energy
  Collisions}}.
\newblock WSP, 2020.

\bibitem{Liu:2020dct}
X.~Liu, F.~Ringer, W.~Vogelsang, and F.~Yuan, {\it {Lepton-jet Correlation in
  Deep Inelastic Scattering}},  \href{http://arxiv.org/abs/2007.12866}{{\tt
  arXiv:2007.12866}}.

\bibitem{Cammarota:2020qcw}
J.~Cammarota, L.~Gamberg, Z.-B. Kang, J.~A. Miller, D.~Pitonyak, A.~Prokudin,
  T.~C. Rogers, and N.~Sato, {\it {The origin of single transverse-spin
  asymmetries in high-energy collisions}},
  \href{http://arxiv.org/abs/2002.08384}{{\tt arXiv:2002.08384}}.

\bibitem{Bacchetta:2020gko}
A.~Bacchetta, F.~Delcarro, C.~Pisano, and M.~Radici, {\it {The
  three-dimensional distribution of quarks in momentum space}},
  \href{http://arxiv.org/abs/2004.14278}{{\tt arXiv:2004.14278}}.

\bibitem{Bacchetta:2019sam}
A.~Bacchetta, V.~Bertone, C.~Bissolotti, G.~Bozzi, F.~Delcarro, F.~Piacenza,
  and M.~Radici, {\it {Transverse-momentum-dependent parton distributions up to
  N$^3$LL from Drell-Yan data}},  \href{http://arxiv.org/abs/1912.07550}{{\tt
  arXiv:1912.07550}}.

\bibitem{Scimemi:2019cmh}
I.~Scimemi and A.~Vladimirov, {\it {Non-perturbative structure of
  semi-inclusive deep-inelastic and Drell-Yan scattering at small transverse
  momentum}},  {\em JHEP} {\bf 06} (2020) 137,
  [\href{http://arxiv.org/abs/1912.06532}{{\tt arXiv:1912.06532}}].

\bibitem{Liu:2018trl}
X.~Liu, F.~Ringer, W.~Vogelsang, and F.~Yuan, {\it {Lepton-jet Correlations in
  Deep Inelastic Scattering at the Electron-Ion Collider}},  {\em Phys. Rev.
  Lett.} {\bf 122} (2019), no.~19 192003,
  [\href{http://arxiv.org/abs/1812.08077}{{\tt arXiv:1812.08077}}].

\bibitem{Chien:2020hzh}
Y.-T. Chien, R.~Rahn, S.~Schrijnder~van Velzen, D.~Y. Shao, W.~J. Waalewijn,
  and B.~Wu, {\it {Azimuthal angle for boson-jet production in the back-to-back
  limit}},  \href{http://arxiv.org/abs/2005.12279}{{\tt arXiv:2005.12279}}.

\bibitem{Chien:2019gyf}
Y.-T. Chien, D.~Y. Shao, and B.~Wu, {\it {Resummation of Boson-Jet Correlation
  at Hadron Colliders}},  {\em JHEP} {\bf 11} (2019) 025,
  [\href{http://arxiv.org/abs/1905.01335}{{\tt arXiv:1905.01335}}].

\bibitem{Fleming:2019pzj}
S.~Fleming, Y.~Makris, and T.~Mehen, {\it {An effective field theory approach
  to quarkonium at small transverse momentum}},  {\em JHEP} {\bf 04} (2020)
  122, [\href{http://arxiv.org/abs/1910.03586}{{\tt arXiv:1910.03586}}].

\bibitem{Buffing:2018ggv}
M.~G.~A. Buffing, Z.-B. Kang, K.~Lee, and X.~Liu, {\it {A transverse momentum
  dependent framework for back-to-back photon+jet production}},
  \href{http://arxiv.org/abs/1812.07549}{{\tt arXiv:1812.07549}}.

\bibitem{Echevarria:2014xaa}
M.~G. Echevarria, A.~Idilbi, Z.-B. Kang, and I.~Vitev, {\it {QCD Evolution of
  the Sivers Asymmetry}},  {\em Phys. Rev. D} {\bf 89} (2014) 074013,
  [\href{http://arxiv.org/abs/1401.5078}{{\tt arXiv:1401.5078}}].

\bibitem{Boer:2015vso}
D.~Boer, C.~Lorcé, C.~Pisano, and J.~Zhou, {\it {The gluon Sivers
  distribution: status and future prospects}},  {\em Adv. High Energy Phys.}
  {\bf 2015} (2015) 371396, [\href{http://arxiv.org/abs/1504.04332}{{\tt
  arXiv:1504.04332}}].

\bibitem{Bacchetta:2006tn}
A.~Bacchetta, M.~Diehl, K.~Goeke, A.~Metz, P.~J. Mulders, and M.~Schlegel, {\it
  {Semi-inclusive deep inelastic scattering at small transverse momentum}},
  {\em JHEP} {\bf 02} (2007) 093,
  [\href{http://arxiv.org/abs/hep-ph/0611265}{{\tt hep-ph/0611265}}].

\bibitem{Kang:2015msa}
Z.-B. Kang, A.~Prokudin, P.~Sun, and F.~Yuan, {\it {Extraction of Quark
  Transversity Distribution and Collins Fragmentation Functions with QCD
  Evolution}},  {\em Phys. Rev.} {\bf D93} (2016), no.~1 014009,
  [\href{http://arxiv.org/abs/1505.05589}{{\tt arXiv:1505.05589}}].

\bibitem{Echevarria:2020qjk}
M.~G. Echevarria, Y.~Makris, and I.~Scimemi, {\it {Quarkonium TMD fragmentation
  functions in NRQCD}},  \href{http://arxiv.org/abs/2007.05547}{{\tt
  arXiv:2007.05547}}.

\bibitem{Callos:2020qtu}
D.~Callos, Z.-B. Kang, and J.~Terry, {\it {Extracting the Transverse Momentum
  Dependent Polarizing Fragmentation Functions}},
  \href{http://arxiv.org/abs/2003.04828}{{\tt arXiv:2003.04828}}.

\bibitem{Arratia:2020nxw}
M.~Arratia, Z.-B. Kang, A.~Prokudin, and F.~Ringer, {\it {Jet-based
  measurements of Sivers and Collins asymmetries at the future Electron-Ion
  Collider}},  \href{http://arxiv.org/abs/2007.07281}{{\tt arXiv:2007.07281}}.

\bibitem{Neill:2016vbi}
D.~Neill, I.~Scimemi, and W.~J. Waalewijn, {\it {Jet axes and universal
  transverse-momentum-dependent fragmentation}},  {\em JHEP} {\bf 04} (2017)
  020, [\href{http://arxiv.org/abs/1612.04817}{{\tt arXiv:1612.04817}}].

\bibitem{Kang:2017btw}
Z.-B. Kang, A.~Prokudin, F.~Ringer, and F.~Yuan, {\it {Collins azimuthal
  asymmetries of hadron production inside jets}},  {\em Phys. Lett. B} {\bf
  774} (2017) 635--642, [\href{http://arxiv.org/abs/1707.00913}{{\tt
  arXiv:1707.00913}}].

\bibitem{Kang:2017glf}
Z.-B. Kang, X.~Liu, F.~Ringer, and H.~Xing, {\it {The transverse momentum
  distribution of hadrons within jets}},  {\em JHEP} {\bf 11} (2017) 068,
  [\href{http://arxiv.org/abs/1705.08443}{{\tt arXiv:1705.08443}}].

\bibitem{Kang:2019ahe}
Z.-B. Kang, K.~Lee, J.~Terry, and H.~Xing, {\it {Jet fragmentation functions
  for $Z$-tagged jets}},  {\em Phys. Lett. B} {\bf 798} (2019) 134978,
  [\href{http://arxiv.org/abs/1906.07187}{{\tt arXiv:1906.07187}}].

\bibitem{Kang:2020xyq}
Z.-B. Kang, K.~Lee, and F.~Zhao, {\it {Polarized jet fragmentation functions}},
   \href{http://arxiv.org/abs/2005.02398}{{\tt arXiv:2005.02398}}.

\bibitem{Gutierrez-Reyes:2019msa}
D.~Gutierrez-Reyes, Y.~Makris, V.~Vaidya, I.~Scimemi, and L.~Zoppi, {\it
  {Probing Transverse-Momentum Distributions With Groomed Jets}},  {\em JHEP}
  {\bf 08} (2019) 161, [\href{http://arxiv.org/abs/1907.05896}{{\tt
  arXiv:1907.05896}}].

\bibitem{Metz:2016swz}
A.~Metz and A.~Vossen, {\it {Parton Fragmentation Functions}},  {\em Prog.
  Part. Nucl. Phys.} {\bf 91} (2016) 136--202,
  [\href{http://arxiv.org/abs/1607.02521}{{\tt arXiv:1607.02521}}].

\bibitem{Collins:1984kg}
J.~C. Collins, D.~E. Soper, and G.~F. Sterman, {\it {Transverse Momentum
  Distribution in Drell-Yan Pair and W and Z Boson Production}},  {\em Nucl.
  Phys.} {\bf B250} (1985) 199--224.

\bibitem{Collins:2011zzd}
J.~Collins, {\it {Foundations of perturbative QCD}},  {\em Camb. Monogr. Part.
  Phys. Nucl. Phys. Cosmol.} {\bf 32} (2011) 1--624.

\bibitem{Ji:2004wu}
X.-d. Ji, J.-p. Ma, and F.~Yuan, {\it {QCD factorization for semi-inclusive
  deep-inelastic scattering at low transverse momentum}},  {\em Phys. Rev. D}
  {\bf 71} (2005) 034005, [\href{http://arxiv.org/abs/hep-ph/0404183}{{\tt
  hep-ph/0404183}}].

\bibitem{Collins:1992kk}
J.~C. Collins, {\it {Fragmentation of transversely polarized quarks probed in
  transverse momentum distributions}},  {\em Nucl. Phys.} {\bf B396} (1993)
  161--182, [\href{http://arxiv.org/abs/hep-ph/9208213}{{\tt hep-ph/9208213}}].

\bibitem{Boer:2003cm}
D.~Boer, P.~J. Mulders, and F.~Pijlman, {\it {Universality of T odd effects in
  single spin and azimuthal asymmetries}},  {\em Nucl. Phys.} {\bf B667} (2003)
  201--241, [\href{http://arxiv.org/abs/hep-ph/0303034}{{\tt hep-ph/0303034}}].

\bibitem{Collins:1981uk}
J.~C. Collins and D.~E. Soper, {\it {Back-To-Back Jets in QCD}},  {\em Nucl.
  Phys.} {\bf B193} (1981) 381. [Erratum: Nucl. Phys.B213,545(1983)].

\bibitem{Boer:1997mf}
D.~Boer, R.~Jakob, and P.~J. Mulders, {\it {Asymmetries in polarized hadron
  production in $e^+e^-$ annihilation up to order 1/Q}},  {\em Nucl. Phys.}
  {\bf B504} (1997) 345--380, [\href{http://arxiv.org/abs/hep-ph/9702281}{{\tt
  hep-ph/9702281}}].

\bibitem{Pitonyak:2013dsu}
D.~Pitonyak, M.~Schlegel, and A.~Metz, {\it {Polarized hadron pair production
  from electron-positron annihilation}},  {\em Phys. Rev. D} {\bf 89} (2014),
  no.~5 054032, [\href{http://arxiv.org/abs/1310.6240}{{\tt arXiv:1310.6240}}].

\bibitem{su:2014wpa}
P.~Sun, J.~Isaacson, C.~P. Yuan, and F.~Yuan, {\it {Nonperturbative functions
  for SIDIS and Drell–Yan processes}},  {\em Int. J. Mod. Phys.} {\bf A33}
  (2018), no.~11 1841006, [\href{http://arxiv.org/abs/1406.3073}{{\tt
  arXiv:1406.3073}}].

\bibitem{Boglione:2016bph}
M.~Boglione, J.~Collins, L.~Gamberg, J.~O. Gonzalez-Hernandez, T.~C. Rogers,
  and N.~Sato, {\it {Kinematics of Current Region Fragmentation in
  Semi-Inclusive Deeply Inelastic Scattering}},  {\em Phys. Lett.} {\bf B766}
  (2017) 245--253, [\href{http://arxiv.org/abs/1611.10329}{{\tt
  arXiv:1611.10329}}].

\bibitem{Hautmann:2020cyp}
F.~Hautmann, I.~Scimemi, and A.~Vladimirov, {\it {Non-perturbative
  contributions to vector-boson transverse momentum spectra in hadronic
  collisions}},  \href{http://arxiv.org/abs/2002.12810}{{\tt
  arXiv:2002.12810}}.

\bibitem{Collins:2004nx}
J.~C. Collins and A.~Metz, {\it {Universality of soft and collinear factors in
  hard-scattering factorization}},  {\em Phys. Rev. Lett.} {\bf 93} (2004)
  252001, [\href{http://arxiv.org/abs/hep-ph/0408249}{{\tt hep-ph/0408249}}].

\bibitem{Bacchetta:2017gcc}
A.~Bacchetta, F.~Delcarro, C.~Pisano, M.~Radici, and A.~Signori, {\it
  {Extraction of partonic transverse momentum distributions from semi-inclusive
  deep-inelastic scattering, Drell-Yan and Z-boson production}},  {\em JHEP}
  {\bf 06} (2017) 081, [\href{http://arxiv.org/abs/1703.10157}{{\tt
  arXiv:1703.10157}}]. [Erratum: JHEP 06, 051 (2019)].

\bibitem{Bauer:2000yr}
C.~W. Bauer, S.~Fleming, D.~Pirjol, and I.~W. Stewart, {\it {An Effective field
  theory for collinear and soft gluons: Heavy to light decays}},  {\em Phys.
  Rev.} {\bf D63} (2001) 114020,
  [\href{http://arxiv.org/abs/hep-ph/0011336}{{\tt hep-ph/0011336}}].

\bibitem{Bauer:2001yt}
C.~W. Bauer, D.~Pirjol, and I.~W. Stewart, {\it {Soft collinear factorization
  in effective field theory}},  {\em Phys. Rev.} {\bf D65} (2002) 054022,
  [\href{http://arxiv.org/abs/hep-ph/0109045}{{\tt hep-ph/0109045}}].

\bibitem{Bauer:2002nz}
C.~W. Bauer, S.~Fleming, D.~Pirjol, I.~Z. Rothstein, and I.~W. Stewart, {\it
  {Hard scattering factorization from effective field theory}},  {\em Phys.
  Rev.} {\bf D66} (2002) 014017,
  [\href{http://arxiv.org/abs/hep-ph/0202088}{{\tt hep-ph/0202088}}].

\bibitem{Beneke:2002ph}
M.~Beneke, A.~P. Chapovsky, M.~Diehl, and T.~Feldmann, {\it {Soft collinear
  effective theory and heavy to light currents beyond leading power}},  {\em
  Nucl. Phys.} {\bf B643} (2002) 431--476,
  [\href{http://arxiv.org/abs/hep-ph/0206152}{{\tt hep-ph/0206152}}].

\bibitem{Becher:2010tm}
T.~Becher and M.~Neubert, {\it {{Drell-Yan} Production at Small $q_T$,
  Transverse Parton Distributions and the Collinear Anomaly}},  {\em Eur. Phys.
  J.} {\bf C71} (2011) 1665, [\href{http://arxiv.org/abs/1007.4005}{{\tt
  arXiv:1007.4005}}].

\bibitem{Chiu:2011qc}
J.-y. Chiu, A.~Jain, D.~Neill, and I.~Z. Rothstein, {\it {The Rapidity
  Renormalization Group}},  {\em Phys. Rev. Lett.} {\bf 108} (2012) 151601,
  [\href{http://arxiv.org/abs/1104.0881}{{\tt arXiv:1104.0881}}].

\bibitem{GarciaEchevarria:2011rb}
M.~G. Echevarria, A.~Idilbi, and I.~Scimemi, {\it {Factorization Theorem For
  Drell-Yan At Low $q_T$ And Transverse Momentum Distributions
  On-The-Light-Cone}},  {\em JHEP} {\bf 07} (2012) 002,
  [\href{http://arxiv.org/abs/1111.4996}{{\tt arXiv:1111.4996}}].

\bibitem{Seidl:2019jei}
{\bf Belle} Collaboration, R.~Seidl et~al., {\it {Transverse momentum dependent
  production cross sections of charged pions, kaons and protons produced in
  inclusive $e^+e^-$ annihilation at $\sqrt{s}=$ 10.58 GeV}},  {\em Phys. Rev.}
  {\bf D99} (2019), no.~11 112006, [\href{http://arxiv.org/abs/1902.01552}{{\tt
  arXiv:1902.01552}}].

\bibitem{Boglione:2017jlh}
M.~Boglione, J.~O. Gonzalez-Hernandez, and R.~Taghavi, {\it {Transverse parton
  momenta in single inclusive hadron production in ${e^ + }{e^ - }$
  annihilation processes}},  {\em Phys. Lett.} {\bf B772} (2017) 78--86,
  [\href{http://arxiv.org/abs/1704.08882}{{\tt arXiv:1704.08882}}].

\bibitem{Soleymaninia:2019jqo}
M.~Soleymaninia and H.~Khanpour, {\it {Transverse momentum dependent of charged
  pion, kaon and proton/antiproton fragmentation functions from $e^+e^-$
  annihilation process}},  \href{http://arxiv.org/abs/1907.12294}{{\tt
  arXiv:1907.12294}}.

\bibitem{Dasgupta:2001sh}
M.~Dasgupta and G.~P. Salam, {\it {Resummation of nonglobal QCD observables}},
  {\em Phys. Lett.} {\bf B512} (2001) 323--330,
  [\href{http://arxiv.org/abs/hep-ph/0104277}{{\tt hep-ph/0104277}}].

\bibitem{Sterman:2004en}
G.~F. Sterman, {\it {Resummations, power corrections and interjet radiation}},
  {\em Acta Phys. Polon.} {\bf B36} (2005) 389--400,
  [\href{http://arxiv.org/abs/hep-ph/0410014}{{\tt hep-ph/0410014}}].

\bibitem{Banfi:2002hw}
A.~Banfi, G.~Marchesini, and G.~Smye, {\it {Away from jet energy flow}},  {\em
  JHEP} {\bf 08} (2002) 006, [\href{http://arxiv.org/abs/hep-ph/0206076}{{\tt
  hep-ph/0206076}}].

\bibitem{Becher:2017nof}
T.~Becher, R.~Rahn, and D.~Y. Shao, {\it {Non-global and rapidity logarithms in
  narrow jet broadening}},  {\em JHEP} {\bf 10} (2017) 030,
  [\href{http://arxiv.org/abs/1708.04516}{{\tt arXiv:1708.04516}}].

\bibitem{Brandt:1964sa}
S.~Brandt, C.~Peyrou, R.~Sosnowski, and A.~Wroblewski, {\it {The Principal axis
  of jets. An Attempt to analyze high-energy collisions as two-body
  processes}},  {\em Phys. Lett.} {\bf 12} (1964) 57--61.

\bibitem{Jain:2011iu}
A.~Jain, M.~Procura, and W.~J. Waalewijn, {\it {Fully-Unintegrated Parton
  Distribution and Fragmentation Functions at Perturbative $k_T$}},  {\em JHEP}
  {\bf 04} (2012) 132, [\href{http://arxiv.org/abs/1110.0839}{{\tt
  arXiv:1110.0839}}].

\bibitem{Lustermans:2016nvk}
G.~Lustermans, W.~J. Waalewijn, and L.~Zeune, {\it {Joint transverse momentum
  and threshold resummation beyond NLL}},  {\em Phys. Lett.} {\bf B762} (2016)
  447--454, [\href{http://arxiv.org/abs/1605.02740}{{\tt arXiv:1605.02740}}].

\bibitem{Dasgupta:2002bw}
M.~Dasgupta and G.~P. Salam, {\it {Accounting for coherence in interjet $E_T$
  flow: A Case study}},  {\em JHEP} {\bf 03} (2002) 017,
  [\href{http://arxiv.org/abs/hep-ph/0203009}{{\tt hep-ph/0203009}}].

\bibitem{Becher:2016omr}
T.~Becher, B.~D. Pecjak, and D.~Y. Shao, {\it {Factorization for the light-jet
  mass and hemisphere soft function}},  {\em JHEP} {\bf 12} (2016) 018,
  [\href{http://arxiv.org/abs/1610.01608}{{\tt arXiv:1610.01608}}].

\bibitem{Becher:2016mmh}
T.~Becher, M.~Neubert, L.~Rothen, and D.~Y. Shao, {\it {Factorization and
  Resummation for Jet Processes}},  {\em JHEP} {\bf 11} (2016) 019,
  [\href{http://arxiv.org/abs/1605.02737}{{\tt arXiv:1605.02737}}]. [Erratum:
  JHEP05,154(2017)].

\bibitem{Moffat:2019pci}
E.~Moffat, T.~Rogers, N.~Sato, and A.~Signori, {\it {Collinear factorization in
  wide-angle hadron pair production in $e^+e^-$ annihilation}},  {\em Phys.
  Rev. D} {\bf 100} (2019), no.~9 094014,
  [\href{http://arxiv.org/abs/1909.02951}{{\tt arXiv:1909.02951}}].

\bibitem{Chiu:2012ir}
J.-Y. Chiu, A.~Jain, D.~Neill, and I.~Z. Rothstein, {\it {A Formalism for the
  Systematic Treatment of Rapidity Logarithms in Quantum Field Theory}},  {\em
  JHEP} {\bf 05} (2012) 084, [\href{http://arxiv.org/abs/1202.0814}{{\tt
  arXiv:1202.0814}}].

\bibitem{Ebert:2019okf}
M.~A. Ebert, I.~W. Stewart, and Y.~Zhao, {\it {Towards Quasi-Transverse
  Momentum Dependent PDFs Computable on the Lattice}},  {\em JHEP} {\bf 09}
  (2019) 037, [\href{http://arxiv.org/abs/1901.03685}{{\tt arXiv:1901.03685}}].

\bibitem{Echevarria:2016scs}
M.~G. Echevarria, I.~Scimemi, and A.~Vladimirov, {\it {Unpolarized Transverse
  Momentum Dependent Parton Distribution and Fragmentation Functions at
  next-to-next-to-leading order}},  {\em JHEP} {\bf 09} (2016) 004,
  [\href{http://arxiv.org/abs/1604.07869}{{\tt arXiv:1604.07869}}].

\bibitem{Luo:2019hmp}
M.-X. Luo, X.~Wang, X.~Xu, L.~L. Yang, T.-Z. Yang, and H.~X. Zhu, {\it
  {Transverse Parton Distribution and Fragmentation Functions at NNLO: the
  Quark Case}},  {\em JHEP} {\bf 10} (2019) 083,
  [\href{http://arxiv.org/abs/1908.03831}{{\tt arXiv:1908.03831}}].

\bibitem{Bauer:2011uc}
C.~W. Bauer, F.~J. Tackmann, J.~R. Walsh, and S.~Zuberi, {\it {Factorization
  and Resummation for Dijet Invariant Mass Spectra}},  {\em Phys. Rev.} {\bf
  D85} (2012) 074006, [\href{http://arxiv.org/abs/1106.6047}{{\tt
  arXiv:1106.6047}}].

\bibitem{Procura:2014cba}
M.~Procura, W.~J. Waalewijn, and L.~Zeune, {\it {Resummation of
  Double-Differential Cross Sections and Fully-Unintegrated Parton Distribution
  Functions}},  {\em JHEP} {\bf 02} (2015) 117,
  [\href{http://arxiv.org/abs/1410.6483}{{\tt arXiv:1410.6483}}].

\bibitem{Sterman:2013nya}
G.~Sterman and M.~Zeng, {\it {Quantifying Comparisons of Threshold
  Resummations}},  {\em JHEP} {\bf 05} (2014) 132,
  [\href{http://arxiv.org/abs/1312.5397}{{\tt arXiv:1312.5397}}].

\bibitem{Becher:2006nr}
T.~Becher and M.~Neubert, {\it {Threshold resummation in momentum space from
  effective field theory}},  {\em Phys. Rev. Lett.} {\bf 97} (2006) 082001,
  [\href{http://arxiv.org/abs/hep-ph/0605050}{{\tt hep-ph/0605050}}].

\bibitem{1808927}
M.~Boglione and A.~Simonelli, {\it {Universality-breaking effects in $e^+e^-$
  hadronic production processes}},  \href{http://arxiv.org/abs/2007.13674}{{\tt
  arXiv:2007.13674}}.

\bibitem{Balsiger:2018ezi}
M.~Balsiger, T.~Becher, and D.~Y. Shao, {\it {Non-global logarithms in jet and
  isolation cone cross sections}},  {\em JHEP} {\bf 08} (2018) 104,
  [\href{http://arxiv.org/abs/1803.07045}{{\tt arXiv:1803.07045}}].

\bibitem{de_Florian_2015}
D.~de~Florian, R.~Sassot, M.~Epele, R.~J. Hernández-Pinto, and M.~Stratmann,
  {\it Parton-to-pion fragmentation reloaded},  {\em Physical Review D} {\bf
  91} (Jan, 2015).

\bibitem{Balsiger:2019tne}
M.~Balsiger, T.~Becher, and D.~Y. Shao, {\it {NLL${'}$ resummation of jet
  mass}},  {\em JHEP} {\bf 04} (2019) 020,
  [\href{http://arxiv.org/abs/1901.09038}{{\tt arXiv:1901.09038}}].

\end{thebibliography}\endgroup
\end{document}